\documentclass[useAMS,usenatbib]{mn2e}

\usepackage{psfig, epsf, epsfig}
\usepackage{graphicx}
\usepackage{amssymb}
\usepackage{pifont} 

\title[P-rich star  formation]{
Phosphorus-rich star formation in giant molecular clouds hosting primordial globular clusters}
\author[K. Bekki \&  T. Tsujimoto]
{Kenji Bekki${}^1$\thanks{E-mail:
kenji.bekki@uwa.edu.au} 
and Takuji Tsujimoto${}^2$ \\
${}^1$ICRAR M468
The University of Western Australia
35 Stirling Hwy, Crawley
Western Australia 6009, Australia \\
${}^2$ 
National Astronomical Observatory of Japan, Mitaka-shi, Tokyo 181-8588, Japan \\
}

\begin{document}

\date{Accepted, Received 2005 February 20; in original form }

\pagerange{\pageref{firstpage}--\pageref{lastpage}} \pubyear{2005}

\maketitle

\label{firstpage}

\begin{abstract}

Recent observational studies have identified phosphorus-rich (P-rich) stars ($[\mathrm{P/Fe}] \ge 1.0$) in both Galactic globular clusters (GCs) including three confirmed cases
and the field population. 
Using analytical models of GC formation, we investigate a scenario in which P-rich stars 
form from pristine gas in GC-forming giant molecular clouds (GMCs) polluted by 
oxygen–neon (ONe) novae within and surrounding the GMCs. Within this framework, we 
model the chemical abundance patterns of P-rich stars using diverse chemical yield 
sets for ONe novae while varying the dilution factor ($F_{\mathrm{dil}}$), defined as 
the ratio of pristine gas mass to total ONe nova ejecta mass. The principal results of our models are as follows:
Across a wide range of $F_{\mathrm{dil}}$, most models display positive correlations 
between $[\mathrm{P/Fe}]$, $[\mathrm{N/Fe}]$, $[\mathrm{O/Fe}]$, $[\mathrm{Al/Fe}]$, $[\mathrm{Si/Fe}]$, $[\mathrm{S/Fe}]$, and $[\mathrm{Si/S}]$. The $[\mathrm{Ce/Fe}]$ 
ratios in P-rich stars can reach values up to $\sim 1$ if the GMCs are 
additionally polluted by interacting low-mass asymptotic giant branch stars. 
P-rich star formation is highly unlikely in environments with a low interaction frequency between GMCs and ONe novae. 
P-rich stars exhibit anomalously low $^{14}\mathrm{N}/^{15}\mathrm{N}$, $^{16}\mathrm{O}/^{17}\mathrm{O}$, and $^{16}\mathrm{O}/^{18}\mathrm{O}$ 
isotopic ratios, offering a clear observational diagnostic to test this scenario. 
Assuming that P-rich Galactic field stars originate from GCs, 
we discuss the origin of their chemical abundance patterns.
We  suggest that the existence of P-rich stars provides direct evidence that GC-forming GMCs were polluted by pre-existing stellar populations.

\end{abstract}

\begin{keywords}
Galaxy:globular clusters --
Galaxy: abundances --
Galaxy:evolution --
Galaxy:formation --
stars: white dwarfs
\end{keywords}

\section{Introduction}

\begin{figure*}
\psfig{file=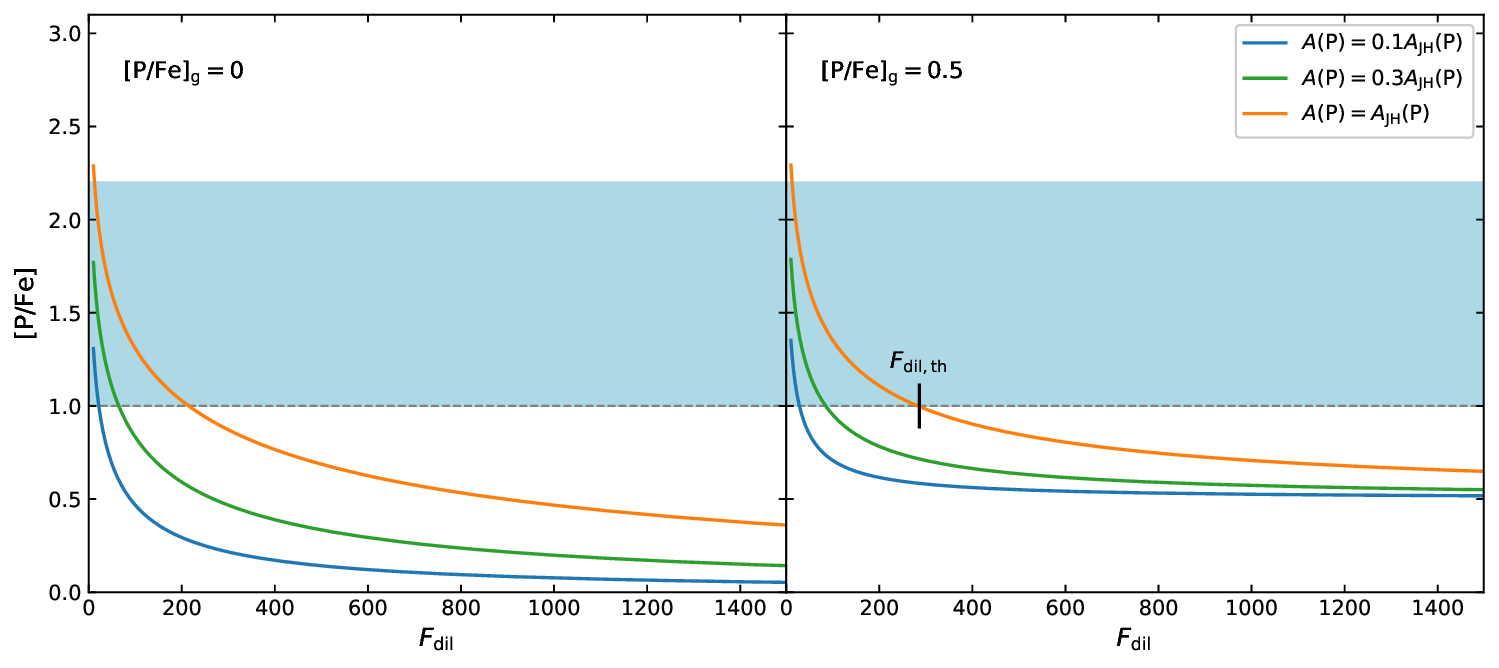,width=18cm}
\caption{
Predicted abundance ratio $\mathrm{[P/Fe]}$ as a function of the gas dilution 
factor $F_{\rm dil}$  for [P/Fe]=0 (left) and 0.5 (right) in the ONe5 models.
The three curves represent different phosphorus yield scaling factors 
relative to standard benchmark predictions 
by JH98, $A_{\rm JH}({\rm P})$: $A({\rm P}) = A_{\rm JH}({\rm P})$ (top), 
$0.3 A_{\rm JH} {\rm (P)} $ (middle), and $0.1 A_{\rm JH}{\rm (P)}$ (bottom).
Stars with [P/Fe]=1 (indicated by a horizontal dotted line) are regarded as P-rich in the present
study.
The horizontal light-blue shaded region 
indicates the range of interest for P-rich stars ($1.0 \le [{\rm P/Fe}] \le 2.2$).
In the right panel, a short vertical black line indicates the threshold 
dilution factor, $F_{\mathrm{dil,th}}$, below which $[\mathrm{P}/\mathrm{Fe}]$ 
exceeds $1.0$.
}
\label{Figure. 1}
\end{figure*}

Recent spectroscopic studies of Galactic stars by the APOGEE survey have discovered 
15 anomalous stars with unusually high phosphorus abundances ($[\mathrm{P}/\mathrm{Fe}] \ge 1$) and 
unusual chemical abundances of several elements, such as Si and Ce (e.g., Masseron et al. 2020a, b, 
M20a, b). Such P-rich stars have also been recently discovered in Galactic GCs 
(e.g., Brauner et al. 2023, B23; Barbuy et al. 2025; Camargo  et al. 2026), 
which suggests that P-enrichment of the intracluster medium by as-yet-unknown stellar sources proceeded efficiently in massive gas clouds forming dense stellar systems. Although the chemical abundance patterns of these P-rich stars have been investigated in detail (e.g., M20a; B23; 
Brauner et al. 2024, B24), the physical origins of their unique chemical abundance patterns are yet to be fully understood. The apparently low fraction of P-rich stars 
among low-metallicity stars  and their
characteristic $[\mathrm{Fe}/\mathrm{H}]$ distribution 
 indicate that some rare astronomical events occurring at a specific epoch of Galaxy formation are associated with P-rich star formation.

Since P-rich stars can potentially form from star-forming gas clouds polluted by stellar objects that eject extremely P-enhanced gas, various ``polluters'' such as asymptotic giant branch (AGB) stars, novae, and particular types of core-collapse supernovae (CCSNe) have been proposed to explain the observed abundances of P-rich stars (M20a). For example, Ritter et al. (2018, R18) 
investigated the nucleosynthesis of massive stars with merging C 
and O shells and found rather high yields of P, Cl, and K in their 3D hydrodynamical 
simulations (see their Fig.~3). If these P-rich ejecta from massive stars can be well-mixed with the ISM to be converted into new stars, $[\mathrm{P}/\mathrm{Fe}]$ of the stars could potentially reach as high as 1.0, depending on dilution factors. Bekki \& Tsujimoto (2024) showed that their chemical evolution models with extremely P-rich ejecta from oxygen-neon (ONe) novae can better reproduce the observed $[\mathrm{Fe}/\mathrm{H}]$--$[\mathrm{P}/\mathrm{Fe}]$ relation of Galactic stars, which implies that P-rich stars might have been formed from such P-rich ejecta of ONe novae. 
Goriely (2022) demonstrated that the observed
chemical abundances of P-rich stars can be well reproduced
by a model with non-thermal nucleosynthesis induced  by 
irradiation processes by energetic photons.

Recent observational studies of Galactic GCs have discovered both P-rich and mildly P-enriched stars ($0.5 \le [\mathrm{P/Fe}] \le 1.0$) and investigated their chemical abundance patterns (e.g., Barbuy et al. 2025; Camargo et al. 2026).
A key result of these observational studies is 
that P-rich stars have been discovered in GCs with very 
small $[\mathrm{Fe}/\mathrm{H}]$ spreads), 
known as ``Type I'' GCs (e.g., Marino et al. 2015). 
This implies that polluters responsible for P-rich star formation should not increase the iron abundances of GC-forming gas clouds, in order to remain consistent with the observed [Fe/H] spreads ($\le 0.05$~dex) in Type I GCs (e.g., Carretta et al. 2009). 
This result therefore suggests that CCSNe with C--O shell merging (e.g., Ritter et al. 2018; Roberti et al. 2025) are unlikely to contribute to the formation of P-rich stars in Type I GCs. It should be noted here, however, that if these massive stars do not explode as CCSNe after enriching star-forming gas through their stellar winds (i.e., collapsing directly to stellar-mass black holes), the above problem of large $[\mathrm{Fe}/\mathrm{H}]$ spreads can be avoided.

\begin{table*}
\centering
\begin{minipage}{180mm}
\caption{Initial chemical abundances of pristine gas in star-forming GMCs
adopted for most models of the present study.[X/Fe] is given in units of dex, and
the solar nitrogen oxygen isotope ratios are adopted in these models.
We investigate the models with the two values of gas-phase
[P/Fe] (${\rm [P/Fe]_g}$=0 and 0.5)  to understand how
${\rm [P/Fe]_g}$ can influence final [P/Fe] of stars.}
\begin{tabular}{lllllllllll}
[Fe/H] & [N/Fe]  &  [O/Fe] & [Mg/Fe] & [Al/Fe] & [Si/Fe] & [P/Fe] & [S/Fe] & $^{14}$N/$^{15}$N & 
$^{16}$O/$^{17}$O  & $^{16}$O/$^{18}$O   \\
-1.0  & 0.0 & 0.76 & 0.4 & 0.0 & 0.25 &  0.5  & 0.25 & $4.4 \times 10^2$ & $2.7 \times 10^3$ 
& $5.0 \times  10^2$  \\
\end{tabular}
\end{minipage}
\end{table*}

Bekki \& McKenzie (2026, BM26) have recently proposed a star-cloud-interaction (``SCI'') 
scenario of GC formation that can accurately reproduce the observed chemical 
abundances of stars in GCs with multiple stellar populations (MPs), 
such as O--Na, Mg--Al, and C--N anticorrelations. In the SCI scenario, pre-existing AGB stars within and around GC-forming GMCs can enrich GMCs with their stellar winds so that new stars with characteristic chemical abundances (``2P'' stars) can be formed (BM26). Like AGB stars, ONe novae can also enrich GMCs close to them if their ejecta are efficiently trapped by GMCs and subsequently well-mixed with the gas. Furthermore, given that ONe nova ejecta are extremely rich in phosphorus 
(e.g., Jos\'{e} \& Hernanz 1998, JH98; Starrfield et al. 1998, 2024; Wanajo et al. 1999;
Psaltis et al. 2025), new stars formed from cold GMC gas polluted by ONe novae are expected to have rather large $[\mathrm{P}/\mathrm{Fe}]$ ($\ge 1$). However, the chemical abundance patterns of P-rich stars formed from the mixture of ONe nova ejecta and GMC gas have yet to be fully investigated by theoretical models of GC formation (BM26). 
It is thus worthwhile to investigate a scenario in which P-rich stars form from the pristine gas of GC-forming GMCs polluted by ONe novae; this scenario is referred to 
as the ``ONe nova scenario'' for convenience.
In this ONe nova scenario, P-rich stars originally formed in low-mass 
GCs are assumed to be tidally stripped during the dynamical disintegration of their 
host clusters due to tidal shock from the Galactic disk and bulge,
eventually becoming field P-rich stars.

The purpose of this paper is to investigate under what physical conditions P-rich stars with $[\mathrm{P}/\mathrm{Fe}] \ge 1$ can form within GC-forming GMCs in the ONe nova scenario using idealized analytical models of GC formation based on the SCI scenario (BM26). 
We particularly investigate the chemical abundances of various elements in P-rich stars, such as $[\mathrm{O}/\mathrm{Fe}]$, $[\mathrm{Si}/\mathrm{Fe}]$, and $[\mathrm{Si}/\mathrm{S}]$, in order to compare the predicted abundances with the corresponding observations in detail. 
We also investigate whether and how chemical enrichment of GMCs by lower-mass AGB stars can explain the observed higher $[\mathrm{Ba}/\mathrm{Fe}]$ (and $[\mathrm{Ce}/\mathrm{Fe}]$) of P-rich stars. 
We provide theoretical predictions on nitrogen and oxygen isotope ratios, because such predictions can serve to test or validate this scenario.
Since P-rich stars are assumed to 
originate from GCs in the present scenario, we compare the predicted
chemical abundances of GC stars with those of P-rich Galactic field stars.

This paper is organized as follows. In \S~2, we present the details of the analytical models used to predict the chemical abundances of P-rich stars in the ONe nova scenario. We describe how the chemical abundances of P-rich stars depend on the adopted $F_{\rm dil}$ and ONe nova types (e.g., ONe1 and ONe2) in \S~3. We discuss the broader physical implications of these findings for observational studies of GCs with P-rich stars and Galactic field stars in \S~4. Finally, we summarize our key conclusions in \S~5. In this paper, we focus exclusively on the chemical abundance patterns of P-rich stars predicted from analytical models, and will discuss the details of P-rich star formation in forming GCs using more sophisticated hydrodynamical simulations in future work.

\begin{figure*}
\psfig{file=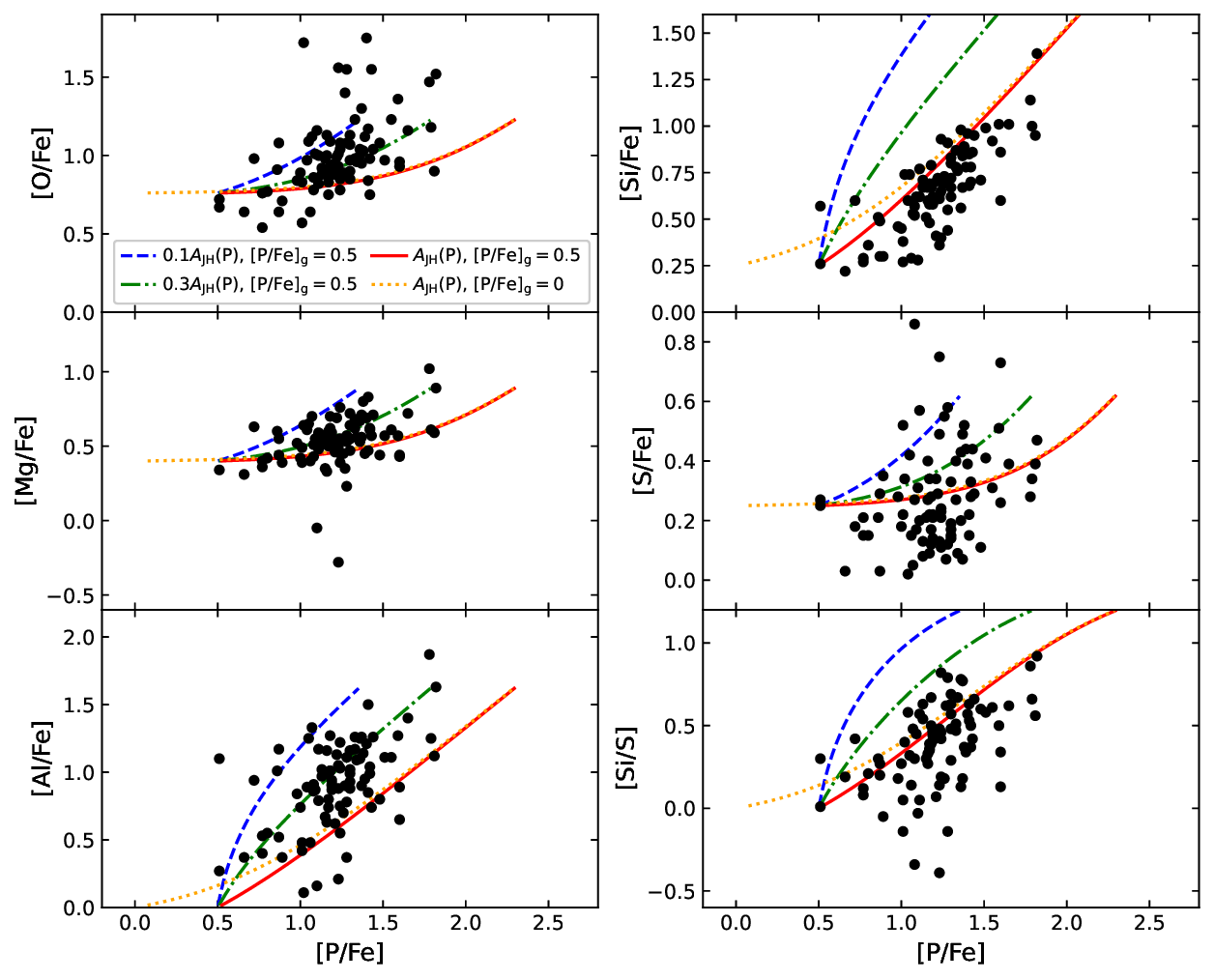,width=18cm}
\caption{
Predicted abundance ratios $[{\rm P/Fe}]$, 
$[{\rm O/Fe}]$, $[{\rm Mg/Fe}]$, $[{\rm Al/Fe}]$, $[{\rm Si/Fe}]$, and $[{\rm S/Fe}]$ 
as a function of the gas dilution factor $F_{\rm dil}$ for the 
four single ONe5 models with different P yield and [P/Fe]$_{\rm g}$:
$A({\rm P}) = 0.1 A_{\rm JH}({\rm P})$ and [P/Fe]$_{\rm g}=0.5$ (blue dashed),
$A({\rm P}) = 0.3 A_{\rm JH}({\rm P})$ and [P/Fe]$_{\rm g}=0.5$ (green dash-dotted),
$A({\rm P}) = A_{\rm JH}({\rm P})$ and [P/Fe]$_{\rm g}=0.5$ (red solid),
and $A({\rm P}) = A_{\rm JH}({\rm P})$ and [P/Fe]$_{\rm g}=0$ (orange dotted).
Observational data points from B23 are also plotted as large black circles.
}
\label{Figure. 2}
\end{figure*}

\section{The model}

In the ONe nova scenario, P-rich stars are assumed to form from 
cold gas within a GMC polluted by ONe novae (see Appendix A for an illustration of the
scenario). 
Because ONe novae cannot synthesize heavy elements such as Ba and La, chemical enrichment 
by ONe novae alone cannot account for the characteristic 
heavy-element abundances observed in P-rich stars, such as high [Ce/Fe] and
[Ba/Fe] (e.g., Brauner et al. 2024, hereafter B24). 
We therefore propose that ejecta from AGB stars with lower masses 
also mix with the pristine gas of the GMC alongside the ONe nova pollution. 
Such synchronized chemical enrichment by both ONe novae and AGB stars is plausible if the GMC is surrounded by stellar populations of varying ages (BM26). 
However, quantifying how ONe novae and AGB stars interact with and enrich surrounding GMCs using hydrodynamical simulations of GMC formation and evolution is beyond the scope of this study.
It should be stressed that while massive AGB stars are 
responsible for the observed anticorrelations between 
light-element abundances (e.g., the C--N anticorrelation), 
lower-mass AGB stars can enhance [Ce/Fe] and [Ba/Fe] in the gas that forms P-rich stars in this scenario.

A key parameter in the scenario is the mass-ratio of 
pristine gas of the GMC ($M_{\rm g}$) 
to ONe nova ejecta ($M_{\rm ej, ONe}$), 
and the parameter is referred to as a ``dilution factor'' :
\begin{equation}
F_{\rm dil} = \frac{ f_{\rm g} M_{\rm gmc} }{ M_{\rm ej, ONe} },
\end{equation}
where $M_{\rm g}$ can be a fraction of the total GMC mass ($M_{\rm gmc}$), because
only a small fraction ($f_{\rm g}$) of the GMC's gas can mix 
with ONe nova ejecta. 
This $f_{\rm g}$ needs to be determined  in our future hydrodynamical simulations of 
GC formation within GMCs.
We mainly investigate the models with $F_{\rm dil}$ ranging from 10
to $10^4$ in the present study.
We use the mass fractions ($A(X)$)  of various elements
($X$=N, O, Mg, Al, Si, S, and P  etc)
for ONe novae predicted by JH98 for the solar metallicity, which investigated
seven ONe nova types, ONe1-7.

These ONe novae exhibit distinct physical properties, 
including the white dwarf mass ($M_{\rm wd}$), 
gas accretion timescale ($t_{\rm acc}$), 
core--envelope mixing fraction, and 
maximum gas temperature ($T_{\rm max}$). 
For instance, $M_{\rm wd}$ (in units of $M_{\odot}$) for models ONe1 
through ONe7 is 1.00, 1.15, 1.15, 1.15, 1.25, 1.35, and 1.35, respectively, 
while their corresponding mixing levels are 50\%, 25\%, 50\%, 75\%, 50\%, 50\%, and 75\%. 
Although the metallicity of GCs containing P-rich stars is much 
lower ($-1.2 < [\mathrm{Fe/H}] < -0.6$) than solar, Jos\'{e} et al. (2007) 
demonstrated that the phosphorus yield fraction is only slightly lower in 
ultra-metal-poor models ($Z = 10^{-6}$) 
than in solar-metallicity models ($Z = 0.02$). 
Given this weak metallicity dependence, along with the lack of 
available ONe nova yield tables spanning $-3 \le [\mathrm{Fe/H}] \le -0.6$, 
we adopt the yields from JH98 as our benchmark reference.

The final abundances ($A_{\rm f}(X)$) of stars can be calculated as follows:
\begin{equation}
A_{\rm f} (X) = \frac{ A_{\rm i}(X)F_{\rm dil}+A(X)}{ 1+ F_{\rm dil} },
\end{equation}
where $A_{\rm i}(X)$ is the initial mass fraction of the element $X$ in a GMC. 
We assumed that these $A_{\rm i}(X)$ are similar to the observed abundances of stars with
[Fe/H]$\approx -1$, because we mainly discuss the chemical abundances of stars with [Fe/H]$-1$.

The adopted initial  abundances are listed in Table 1, which are consistent with
average chemical abundances of the Galactic halo stars with [Fe/H]$\approx -1$ in the
SAGA data base (Suda et al. 2008): 
we use a slightly lower [N/Fe] ($=0$) in the present models to discuss the origin
of [N/Fe] of 1P and 2P stars in GCs.
The initial [P/Fe] in pristine gas, which is denoted as ${\rm [P/Fe]_g}$, is a key
parameter that can control [P/Fe] of stars in the present study.
Accordingly, we investigate the models with [P/Fe]=0 and 0.5 to cover the observed
range of [P/Fe] of metal-poor Galactic stars.
The P abundances of ONe novae predicted from JH98 are referred to as $A_{\rm JH} {\rm (P)}$,
and we investigate $A{\rm (P)}=[0.1, 0.3, 1.0]A_{\rm JH} {\rm (P)}$ due to
the possible uncertainty in P yields from ONe novae with low [Fe/H].
J07 and Psatis et al. (2025) investigated P yields for extremely metal-poor
ONe novae and found that P yields  do not depend so strongly on metallicities: only
a factor of $\approx 2$ lower in $Z=10^{-4} Z_{\odot}$)
compared with $Z=Z_{\odot}$
in Fig. 1 by J07.
These results imply that P yields for [Fe/H]$\approx -1$ may not be so different from those
for [Fe/H]=0.  Nevertheless, we investigate the above three types of P yields in the present
study.

\begin{figure*}
\psfig{file=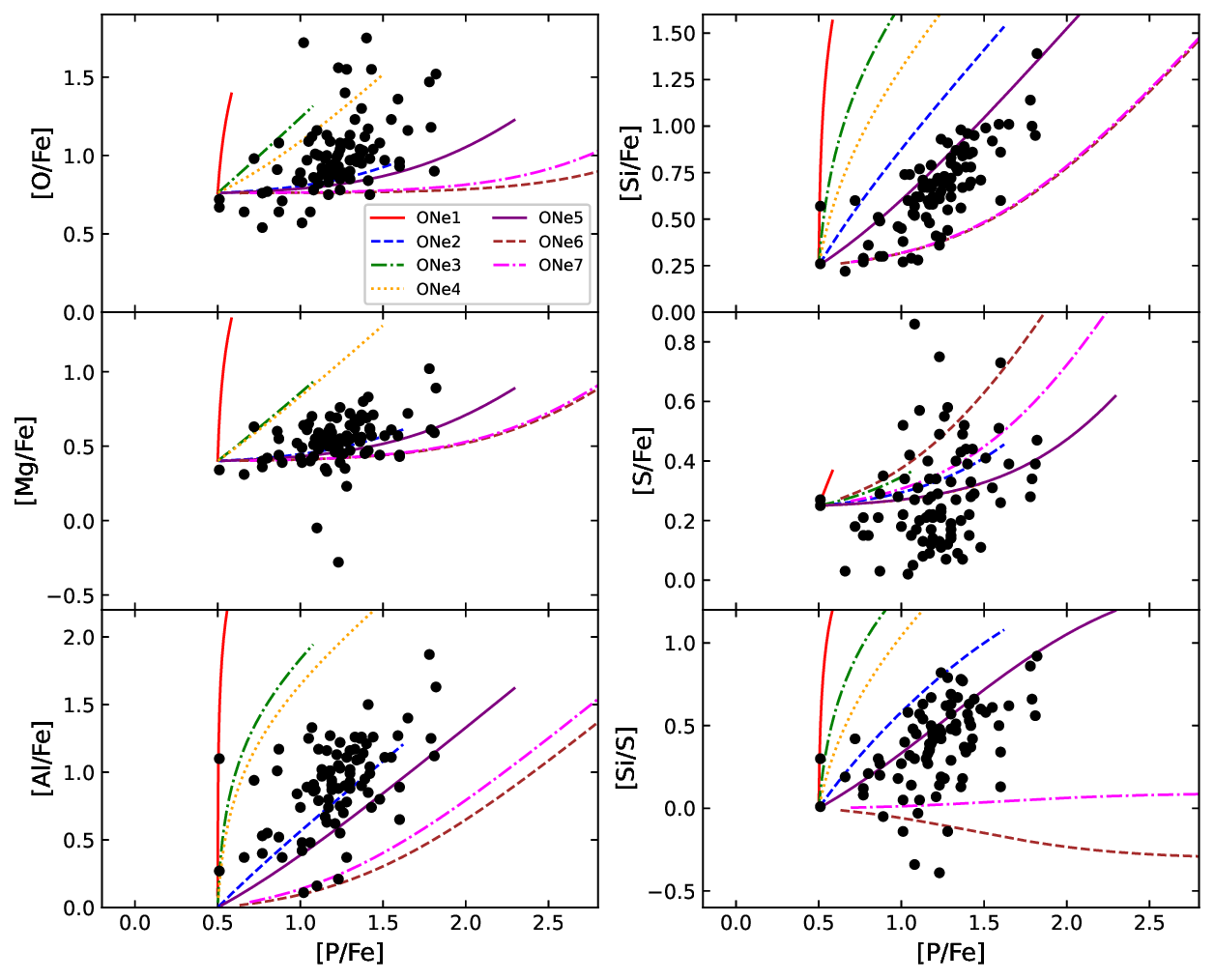,width=18cm}
\caption{
Same as Fig. 2, but for different ONe models:
red solid (ONe1),
blue dashed (ONe2),
green dash-dot (ONe3),
orange dotted (ONe4),
purple solid (ONe5),
brown dashed (ONe6),
and magenta dash-dot (ONe7).
}
\label{Figure. 3}
\end{figure*}

The phosphorus abundances of ONe novae predicted by JH98 are referred to as $A_{\mathrm{JH}}(\mathrm{P})$. To account for potential uncertainties in P yields from low-metallicity ONe novae, 
we test $A(\mathrm{P}) = [0.1, 0.3, 1.0] \times A_{\mathrm{JH}}(\mathrm{P})$. 
J07 and Psatis et al. (2025) investigated P yields 
for extremely metal-poor ONe novae and found that they do not depend 
strongly on metallicity; specifically, the yields are only a factor of $\approx 2$ lower at $Z = 10^{-4}\,Z_{\odot}$ compared to $Z = Z_{\odot}$ in Fig.~1 by J07. 
These findings imply that P yields at $[\mathrm{Fe}/\mathrm{H}] \approx -1$ are likely similar to those at $[\mathrm{Fe}/\mathrm{H}] = 0$. Nevertheless, we explore all three yield cases in the present work.

We investigate both ``single'' and ``composite'' models in the present study. Only one ONe nova type
(e.g., ONe1 only) is assumed to interact with GMCs in the single models, whereas multiple types 
(e.g., ONe1, ONe2, and ONe3)
assumed to  interact with
GMCs in the composite ones:
\begin{equation}
A(X)=\frac{1}{ m_{\rm ej,t} }  \sum_{\it j=1}^{\rm N} P_{\rm j} A_{\it j}(X)m_{\rm ej, \it j},
\end{equation}
where $N$ is the number of ONe nova types, $A_{\rm j}$, 
$P_{\it j}$,
and $m_{\rm ej, \it j}$ are
the yield,
the fraction, and the total ejecta  mass
for $j$-th ONe nova, respectively.
$m_{\rm ej, t}$ is 
is the total ejecta mass from seven ONe nova types and estimated as follows:
\begin{equation}
m_{\rm ej,t} =\sum_{\it j=1}^{\rm N} P_{\rm j} m_{\rm ej, \it j}.
\end{equation}
Using a random number generator, we allocate $P_j$ for $j$-th ONe nova type:
$\sum_{\it j=1}^{\rm N} P_{\rm j}$ should be 1.
We mainly investigate the composite models with $N=3$ and 7 in the present study.

We define the mass fraction of P-rich stars in a globular cluster (GC) as follows:
\begin{equation}
M_{\rm P-rich}=F_{\rm P-rich} = \frac{M_{\rm P-rich}}{M_{\rm gc}},
\end{equation}
where $M_{\rm P-rich}$ and $M_{\rm gc}$ are the total 
mass of P-rich stars and the total mass of the GC, respectively. Since $F_{\rm dil} \gg 1$, $M_{\rm P-rich}$ can be expressed as a function of the star formation efficiency ($\epsilon$), $M_{\rm gc}$, $F_{\rm dil}$, and the ONe ejecta mass ($M_{\rm ej, ONe}$):
\begin{equation}
F_{\rm P-rich} M_{\rm gc} = \epsilon M_{\rm ej, ONe} F_{\rm dil}.
\end{equation}
Although the exact range of $F_{\rm dil}$ cannot be derived from first principles in this 
study, we can infer a plausible range based on the numerical simulation results 
reported in BM26. As discussed in Section 4.1, 
a massive GMC with $M_{\rm gmc} = 10^6~{\rm M}_{\odot}$ can potentially 
interact with $M_{\rm ej, ONe} \approx 10^{-5} M_{\rm gmc} \approx 10~{\rm M}_{\odot}$. Consequently, a value of $F_{\rm dil} = 10^4$ is obtained for $M_{\rm gc} = 10^5~{\rm M}_{\odot}$, $\epsilon = 0.1$, and $F_{\rm P-rich} = 0.1$. 
If $F_{\rm P-rich}$ can be substantially lower than $0.1$, $F_{\rm dil}$ is 
expected to be well below $10^4$ in many cases. For example, $F_{\rm dil} = 10$ corresponds to a low-mass GC with $M_{\rm gc} = 10^4~{\rm M}_{\odot}$, $\epsilon = 0.1$, and $F_{\rm P-rich} = 0.001$. Although $F_{\rm dil} \approx 10^5$ could be realized for lower star formation efficiencies (e.g., $\epsilon \approx 0.01$), 
we consider $10 \le F_{\rm dil} \le 10^4$ to be a reasonable range 
within the framework of the ONe nova scenario.
The above discussion of $F_{\rm dil}$ is purely qualitative, 
as $M_{\rm ej, ONe}$ can vary significantly among GMCs with 
different $M_{\rm gmc}$. 
Hydrodynamical simulations of GC formation are required to investigate the 
dilution processes of ONe nova ejecta within GMCs and enable a more quantitative assessment of the plausible range of $F_{\rm dil}$.

In the ONe nova scenario,  AGB stars with lower masses ([3-4] ${\rm M}_{\odot}$) 
 can also interact with star-forming GMCs so that
the new stars can have the enhanced abundances in  Ce, Sr, and Ba etc. We here investigate
[Ce/Fe] using yields tables for low-metallicity AGB stars predicted by Fishlock et al. (2014)
and Lugaro et al. (2012).
The key parameter for the final [Ce/Fe] in the scenario is the dilution factor ratio
($R_{\rm dil}$) of ONe novae to AGB stars;
\begin{equation}
R_{\rm dil} =\frac{ F_{\rm dil, agb} }{ F_{\rm dil} },
\end{equation}
where $F_{\rm dil, agb}$ is the dilution factor for AGB stars.
We consider that this $R_{\rm dil}$ can be quite different between different GMCs,
and accordingly regard it as a free parameter.
We mainly investigate the models with $R_{\rm dil}=0.1$, 1, and 10 to illustrate
how this parameter can determine the final [Ce/Fe] of P-rich stars,
though the lower $R_{\rm dil}$ ($<0.1$)  might be possible.
We will need to investigate a possible range of $R_{\rm dil}$ in our future studies by using
hydrodynamical simulations of GMC formation and evolution surrounded by ONe novae and AGB stars.

We mainly investigate [O/Fe], [Mg/Fe], [Si/Fe], and [P/Fe] of P-rich stars,
because recent observational studies (e.g., B23) have measured these abundances and found 
significant enhancement in these.  We also try to find correlations of [P/Fe] with
[O/Fe], [Mg/Fe], [Si/Fe], and [Si/S] and thereby compare them with the corresponding observations.
Since P-rich stars in M20a have [P/Fe] ranging from 1.0 to 2.0,  we invesgiate 
the possible ranges of chemical abundances
for various elements including oxygen and nitrogen abundances for these P-rich stars
with $1 \le {\rm [P/Fe]} \le 2.2$ in the present study.
We investigate the threshold dilution factor, $F_{\mathrm{dil,th}}$, below which $[\mathrm{P}/\mathrm{Fe}] \ge 1.0$, defining the condition for P-rich star formation in each model as:
\begin{equation}
F_{\mathrm{dil}} \le F_{\mathrm{dil,th}}.
\end{equation}
This threshold $F_{\mathrm{dil,th}}$ varies significantly across models with 
different ONe nova types and yield assumptions due to large variations in their predicted phosphorus yields. Fig.~1 illustrates how $F_{\mathrm{dil}}$ determines the 
final stellar $[\mathrm{P}/\mathrm{Fe}]$ in ONe5 models for various adopted P yields.
Clearly, $F_{\mathrm{dil,th}}$ must be larger for higher 
P yields owing to the greater chemical enrichment by ONe novae.
Although the initial gas-phase abundance ratio, $[\mathrm{P}/\mathrm{Fe}]_{\mathrm{g}}$, 
can influence $F_{\mathrm{dil,th}}$, 
its effect is significantly less pronounced than 
that of the ONe nova types and yields. Therefore, in this study, 
we focus primarily on how the results depend on the ONe nova types and yields.

\begin{figure*}
\psfig{file=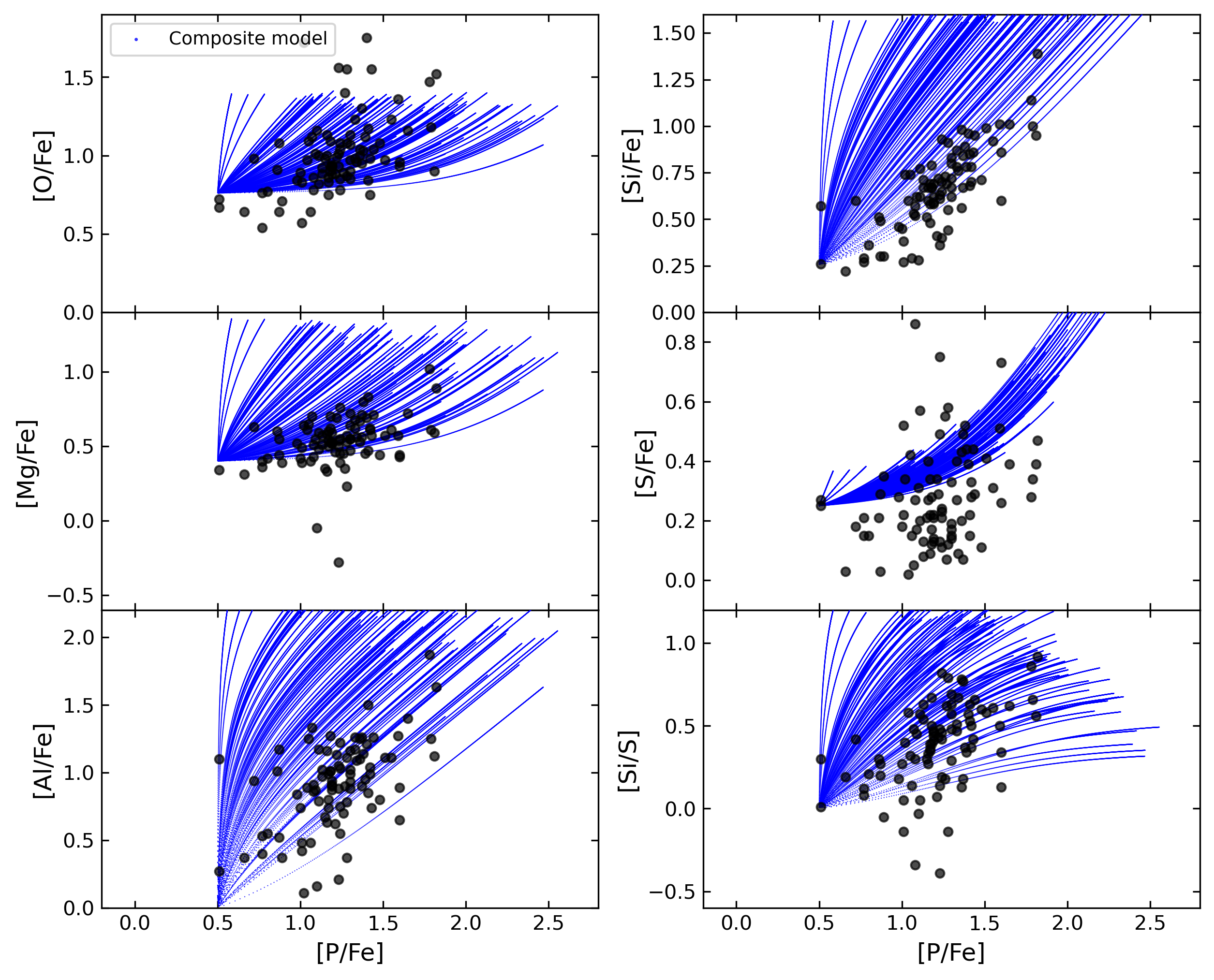,width=18cm}
\caption{
Same as Fig. 2, but for composite models combining different ONe nova types 
and values of $F_{\mathrm{dil}}$. 
Small blue dots denote individual model predictions for the chemical abundances.
}
\label{Figure. 4}
\end{figure*}

\section{Results}

\subsection{Single models}

Fig.~2 illustrates the dependence of the predicted abundance ratios ($[{\rm P/Fe}]$, $[{\rm O/Fe}]$, $[{\rm Mg/Fe}]$, $[{\rm Al/Fe}]$, $[{\rm Si/Fe}]$, $[{\rm S/Fe}]$, and $[{\rm Si/S}]$) on $[{\rm P/Fe}]$ across the four single ONe5 models. 
These models were selected because they reproduce the observed 
abundance patterns of B23 better than other ONe models. For all models, these abundance ratios increase monotonically as $F_{\rm dil}$ decreases, leading to positive 
correlations between $[{\rm P/Fe}]$ and $[{\rm O/Fe}]$, $[{\rm Mg/Fe}]$, $[{\rm Al/Fe}]$, $[{\rm Si/Fe}]$, and $[{\rm S/Fe}]$. The dependence of $[{\rm Si/Fe}]$ on $[{\rm P/Fe}]$ is 
significantly stronger than that of $[{\rm S/Fe}]$ in these ONe nova models; consequently, $[{\rm Si/S}]$ also depends on $[{\rm P/Fe}]$ such that $[{\rm Si/S}]$ is 
higher at higher $[{\rm P/Fe}]$. Clearly, P-rich stars ($[{\rm P/Fe}] \ge 1$) show substantial enhancements in $\alpha$-elements and aluminum, which is at least qualitatively 
consistent with observations. The observed relation between $[{\rm S/Fe}]$ and $[{\rm P/Fe}]$ shows larger scatter, 
likely reflecting the higher dispersion of $[{\rm S/Fe}]$ in the pristine gas of 
star-forming GMCs.

Although the model with $A({\rm P}) = 0.3 A_{\rm JH}({\rm P})$ provides a 
better fit to the observed $[{\rm O/Fe}]-[{\rm P/Fe}]$, $[{\rm Mg/Fe}]-[{\rm P/Fe}]$, $[{\rm S/Fe}]-[{\rm P/Fe}]$, and $[{\rm Al/Fe}]-[{\rm P/Fe}]$ relations compared to those with $A({\rm P}) = 0.1 A_{\rm JH}({\rm P})$ and $A({\rm P}) = A_{\rm JH}({\rm P})$, 
the model with $A({\rm P}) = A_{\rm JH}({\rm P})$ best 
reproduces the $[{\rm Si/Fe}]-[{\rm P/Fe}]$ and $[{\rm Si/S}]-[{\rm P/Fe}]$ relations. Accordingly, it is difficult for the present study to place strong 
constraints on the theoretical yields of ONe5 novae in these models. 
Clearly, models with $A({\rm P}) = 0.1 A_{\rm JH}({\rm P})$ do not yield better fits to the observational data, implying that the P yields for low-metallicity ONe novae 
should not be too small to reproduce the observations. 
Indeed, J07 demonstrated that P yields do not depend strongly on metallicity, although they investigated only models with solar metallicity ($Z = Z_{\odot}$), $Z = 10^{-4} Z_{\odot}$, and $Z = 10^{-5} Z_{\odot}$. Given that yields for ONe novae with $Z \approx 10^{-1} Z_{\odot}$ are currently unavailable, future theoretical investigations into the metallicity dependence of ONe nova yields are highly desirable to assess the validity of the ONe nova scenario.

As shown in Fig.~2, these ONe5 models appear to 
overpredict $[{\rm Si/S}]$ for a given $[{\rm P/Fe}]$ across different P 
yields and $[{\rm P/Fe}]_{\rm g}$. This could simply be due to adopting 
fixed single values for the initial gas-phase Si and S abundances in the pristine gas of GMCs ($0.25$ and $0.25$, respectively). 
If a lower value (e.g., $0$) is adopted for these gaseous abundances, the 
observational data can be better fitted by the models. Therefore, we consider that this apparent overprediction of $[{\rm Si/S}]$ in these models is not a serious problem for the ONe nova 
scenario. The observed larger scatter in $[{\rm Si/S}]$ for a given $[{\rm P/Fe}]$ could result from a higher intrinsic dispersion of $[{\rm S/Fe}]$ in the pristine gas of GMCs. 
Although the observed range of $[{\rm Si/S}]$ in P-rich stars cannot be fully explained by single ONe5 models alone, the present study demonstrates, for the first time, that the ONe nova scenario can reproduce the observed positive correlation between $[{\rm Si/S}]$ and $[{\rm P/Fe}]$, which is a key characteristic of P-rich stars.

As shown in the observed distribution of P-rich stars on the $[{\rm Mg/Fe}]-[{\rm P/Fe}]$ plane, 
only a few stars exhibit low $[{\rm Mg/Fe}]$ ($<0$, i.e., Mg-depleted), 
similar to second-generation (2P) stars in metal-poor globular clusters (GCs). 
This implies that only $\approx 3\%$ of P-rich stars are likely to originate from 2P GC populations. Since chemical enrichment by ONe novae can increase $[{\rm Mg/Fe}]$ 
by at most $0.3~{\rm dex}$ per $1~{\rm dex}$ increase in $[{\rm P/Fe}]$, the cold gas from which Mg-depleted P-rich stars formed might have had an even lower initial $[{\rm Mg/Fe}]$. 
It should be stressed that there are no P-rich stars with O depletion in B23
, which is otherwise a key 
hallmark of 2P stars in GCs. 
Also it is confirmed by the authors of this paper that no O depletion can be found
in P-rich stars for M4, NGC 6569, and NGC 6539.
These observed O and Mg abundances accordingly suggest that most P-rich stars belong to 1P GC populations. 
If this is the case, ONe novae must occur well before GMCs 
are enriched by the stellar winds of massive AGB stars to form P-rich stars without O and Mg depletion. The formation of Mg-depleted P-rich stars may represent a relatively rare case where GMCs are polluted by ONe novae after being polluted by massive AGB stars. 
Given that O and Mg depletion are absent in most P-rich stars, 
the observed enhancement in Al abundances is likely driven by chemical enrichment from ONe novae rather than from massive AGB stars.

The small fraction of P-rich stars that likely originate from second-generation (2P) stars in GCs
 also implies that P-rich stars form in GMCs
 that host only low-mass GCs. The fraction of 2P stars ($F_{\rm 2P}$) in 
GCs is observed to depend on GC mass, such that $F_{\rm 2P}$ is smaller in less massive GCs 
(e.g., Milone \& Marino 2022). Such low-mass GCs are highly 
susceptible to disintegration through long-term dynamical relaxation processes 
and tidal disruption by the Galactic gravitational field. Therefore, 
if the observed field P-rich stars originated from defunct low-mass GCs, the small observed fraction of P-rich stars with Mg depletion can be naturally explained within the ONe nova scenario. Due to the small sample size of P-rich stars identified in GCs to date, it remains observationally unclear whether P-rich stars in GCs are preferentially 1P or 2P stars.

Fig.~3 shows the predicted abundance ratios as a function of $[{\rm P/Fe}]$ 
for seven single ONe nova models (${\rm ONe1}$--${\rm ONe7}$) with varying $F_{\rm dil}$, 
assuming $[{\rm P/Fe}]_{\rm g} = 0.5$ and $A({\rm P}) = A_{\rm JH}({\rm P})$. Clearly, 
the ONe1, ONe3, and ONe4 models predict excessively high abundances of $\alpha$-elements and Al, 
suggesting that these models are disfavored in the ONe nova scenario. Although the ONe2 model provides a better fit to the observed abundances of $\alpha$-elements and Al, it overpredicts $[{\rm Si/Fe}]$ across the adopted range of $F_{\rm dil}$. Conversely, the ONe6 and ONe7 models, which feature large P yields, underpredict these chemical abundances (with the exception of Si), indicating that they are also not optimal models within this framework. 
Although the ONe1, 2, 3, 4, 6, and 7 models are inferior to the ONe5 models in self-consistently reproducing the observational dataset as a whole, they can reproduce certain observational features that cannot be explained by single ONe5 models alone. 
This implies that individual P-rich stars may form in GMCs polluted by different types of ONe novae.

Physically, there is no fundamental reason why only ONe5 novae should 
preferentially pollute star-forming GMCs surrounded by 
field stellar populations containing diverse ONe nova 
progenitor types. Therefore, P-rich star formation in GMCs 
enriched by various types of ONe novae appears to be both realistic and reasonable. 
It is plausible and perhaps even more realistic that a P-rich star forms from 
cold GMC gas polluted by multiple types of ONe novae with varying mixing proportions 
in their ejecta. 
If this is the case, the observed wide dispersion in $\alpha$-element and 
Al abundances for a given $[{\rm P/Fe}]$ can be naturally accounted for within the ONe nova scenario. 
Consequently, investigating whether such multi-type ONe nova pollution in GMCs is physically feasible using hydrodynamical simulations of GC formation will be the subject of our future work.

\begin{figure}
\psfig{file=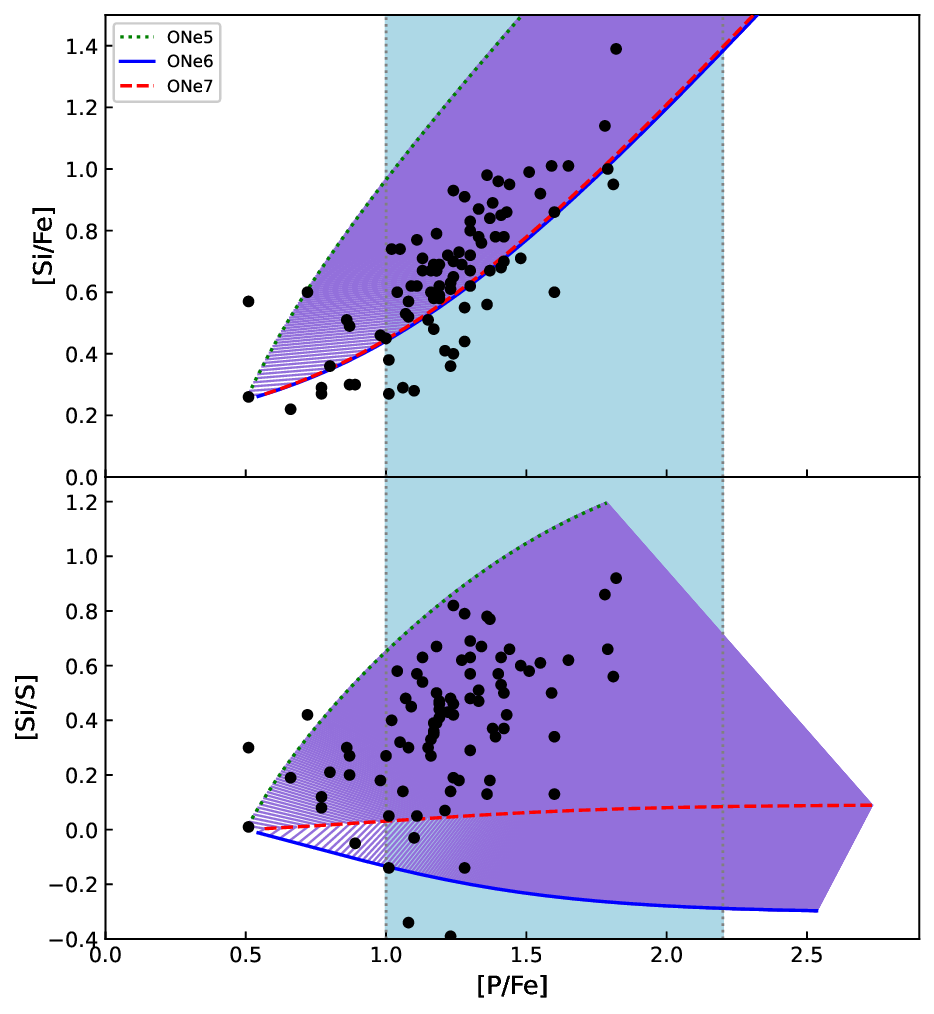,width=8.5cm}
\caption{
Predicted abundance ratios $[{\rm Si/Fe}]$ (top panels) and $[{\rm Si/S}]$ (bottom panels) 
as a function of $[{\rm P/Fe}]$ for the composite models combining ONe5 (greed dotted), 
6 (blue solid), and 7 (red dashed)  models
for $A({\rm P}) = 0.3 A_{\rm JH}({\rm P})$.
The vertical light-blue shaded region 
indicates the range of interest for P-rich stars ($1.0 \le [{\rm P/Fe}] \le 2.2$).
The light-purple shaded region indicates the ranges of $[\mathrm{P}/\mathrm{Fe}]$,
 $[\mathrm{Si}/\mathrm{Fe}]$, and $[\mathrm{S}/\mathrm{Fe}]$ predicted by the 
composite models of the three ONe types.
}
\label{Figure. 5}
\end{figure}

\begin{figure*}
\psfig{file=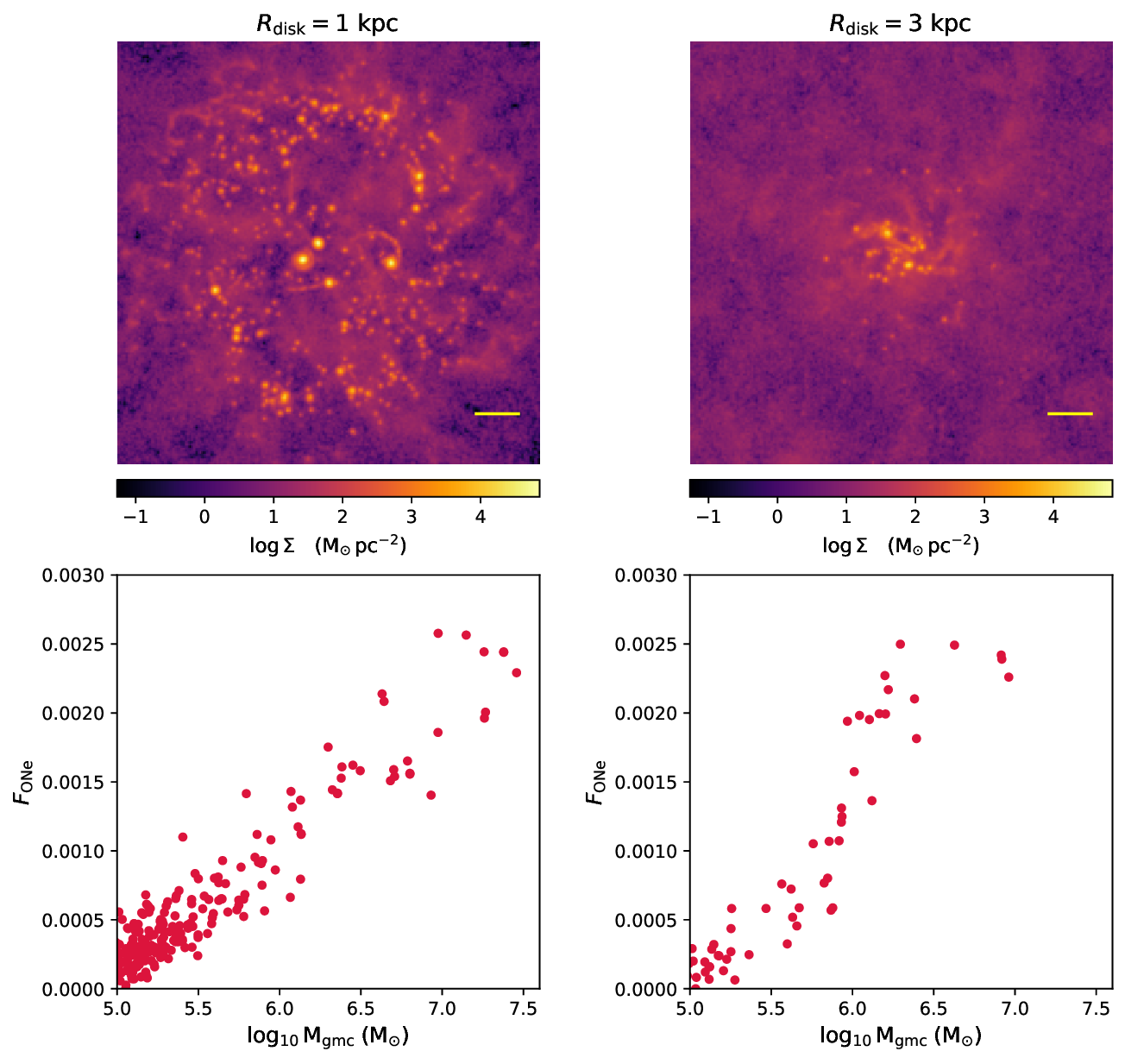,width=18cm}
\caption{
Surface gas  mass density, $\log \Sigma$ (${\rm M}_{\odot}~{\rm pc}^{-2}$) in dwarf galaxies (upper)
and the predicted mass fraction of ONe novae, $F_{\rm ONe}$, 
within star-forming GMCs 
as a function of the GMC mass, $\log_{10} M_{\rm gmc}$ (${\rm M}_{\odot}$) (lower). 
Each filled red circle in the lower panel corresponds to 
an individual GMC formed in this simulation.
The results are shown for two dwarf
galaxy models with different disk sizes, 
indicated by $R_{\rm disk} = 1~{\rm kpc}$ and $R_{\rm disk} = 3~{\rm kpc}$.
The scale bar in the bottom-right corner of each panel represents 200 pc.}
\label{Figure. 6}
\end{figure*}

\subsection{Composite models}

Although single ONe models demonstrate that the ONe5 model best 
explains the observed properties of P-rich stars, GMCs are realistically 
polluted by various types of ONe novae. Fig.~4 shows the predicted 
abundance ratios ($[{\rm O/Fe}]$, $[{\rm Mg/Fe}]$, $[{\rm Al/Fe}]$, $[{\rm Si/Fe}]$, $[{\rm S/Fe}]$, and $[{\rm Si/S}]$) as a function of $[{\rm P/Fe}]$ for composite 
ONe nova models across different values of $F_{\rm dil}$. 
In the present study, each composite model comprises $100$ distinct 
combinations of all ONe nova types for $1000$ different values 
of $F_{\rm dil}$. Unlike single ONe models, composite models yield a range of $[{\rm P/Fe}]$ values at any given $F_{\rm dil}$ due to varying fractional contributions from different ONe nova subtypes. Consequently, the predicted chemical abundances of $\alpha$-elements, Al, and $[{\rm Si/S}]$ display large dispersions for a given $[{\rm P/Fe}]$, allowing these composite models to cover wider observed abundance ranges in P-rich stars.

Although these composite models reproduce the observed chemical abundances 
better than single-type models, 
several discrepancies remain. 
First, several O-rich stars with $[{\rm O/Fe}] > 1.4$ cannot be adequately 
reproduced by any combination of ONe nova types; however, this might simply 
reflect an initially high $[{\rm O/Fe}]$ in the pristine gas of the host GMCs
(There are no stars with [Fe/H]$>-2$ and [O/Fe]$>1.5$ in the SAGE data base, however).
Second, the predicted range of $[{\rm S/Fe}]$ for a given $[{\rm P/Fe}]$ is significantly narrower than observed, which could likewise stem from a larger intrinsic 
dispersion of $[{\rm S/Fe}]$ in GMCs. 
Third, no P-rich stars with $[{\rm P/Fe}]$ as high as $2.5$ have been 
observed, whereas such extreme abundances are predicted by composite 
models with smaller $F_{\rm dil}$. One possible solution to this apparent 
inconsistency is that gaseous ejecta from ONe6 and ONe7 novae—which are 
responsible for $[{\rm P/Fe}] \approx 2.5$ in the composite models—cannot 
be retained within GMCs due to high velocity of ejecta. Alternatively, 
ONe nova explosions arising from binary systems with 
white dwarf masses of $m_{\rm WD}=1.35~{\rm M}_{\odot}$ may not occur within the typical lifetimes of GMCs (i.e., a few times $10^7~{\rm yr}$). 
Regardless of which mechanism resolves the third discrepancy, we consider this issue not to pose a fundamental challenge to the scenario.

In these composite models, the fractional contributions of the seven 
ONe nova types to GMC chemical enrichment are allocated randomly. 
This uniform random allocation may be simplistic and idealized, 
as stars with lower initial masses are more abundant under a standard stellar 
initial mass function (IMF). 
Consequently, ONe1 novae (originating from lower-mass white dwarfs, $m_{\mathrm{WD}}$) 
should interact with GMCs more frequently than ONe7 novae (from higher-mass $m_{\mathrm{WD}}$). 
Because the relative contributions of various ONe nova types 
depend strongly on the adopted IMF—which dictates the abundance of intermediate-mass 
stars across different masses—future studies 
will need to investigate more realistic composite models tailored to specific IMFs
to consider possible IMF variations within GMCs.

Fig.~5 presents $[{\rm Si/Fe}]$ and $[{\rm Si/S}]$ as a function of $[{\rm P/Fe}]$ for 
composite ONe nova models incorporating only ONe5, ONe6, and ONe7 (excluding ONe1--4) with $A({\rm P}) = 0.3 A_{\rm JH}({\rm P})$ (note that $A({\rm P}) = A_{\rm JH}({\rm P})$ 
was adopted in Fig.~4). 
The light-blue shaded region highlights the observed range of 
phosphorus enrichment ($1.0 \le [{\rm P/Fe}] \le 2.2$) in P-rich stars, 
whereas the light-purple shaded region indicates the domain covered by these composite models. 
Owing to the smaller adopted P yields and the inclusion of only ONe5--7, 
these models better reproduce the observed ranges of $[{\rm Si/Fe}]$ and $[{\rm Si/S}]$, 
both of which define key observational characteristics of P-rich stars. 
Although the predicted $[{\rm Si/Fe}]$ ratios remain systematically higher than observed, 
these more consistent results imply that P-rich stars formed from GMCs polluted 
primarily by ONe novae with higher-mass white dwarfs. 
The observed range of $[{\rm P/Fe}]$ in P-rich stars 
suggests that $F_{\rm dil}$ must span from $10$ to $769$, 
with a mean value of $53$ in these composite models. 
This requirement for a certain range of
$F_{\rm dil}$ values is consistent with the predictions of single ONe nova models (see Fig.~1).

The results shown in Figs.~2--5 demonstrate that: (i) the ONe nova scenario is 
promising for self-consistently reproducing the various observed chemical abundances 
of P-rich stars, and (ii) GMCs must be polluted by ONe novae with a
certain range of $F_{\mathrm{dil}}$ values ($F_{\rm dil} < 10^3$). 
This requirement for  $F_{\mathrm{dil}}$ implies that 
gaseous ejecta from ONe novae interact with only a small 
fraction of the GMC's pristine gas, likely driven by specific 
physical mechanisms. Consequently, two key questions arise: 
(1) how do ONe novae interact so frequently with GMCs, and 
(2) why is a specific range of $F_{\mathrm{dil}}$ 
required in these interactions between GMCs and ONe novae? 
Although full hydrodynamical simulations of GMCs interacting with ONe novae in galaxies are ultimately required to resolve these issues, we address them using idealized analytical models in the following section.

\section{Discussion}

Under the ONe nova scenario, which assumes that phosphorus-rich (P-rich) Galactic 
field stars originate from globular clusters (GCs), their age,
 metallicity, and spatial distributions should mirror those of GCs disrupted by 
Galactic dynamics (e.g., disk shocking). Consequently, it is 
expected that P-rich field stars in this scenario are old and 
concentrated within the inner Galaxy, the halo, and GCs. 
Below, we address several key issues concerning Galactic P-rich stars in light of recent observational data.

\subsection{Probabilities of GMCs interacting with ONe novae}

In the ONe nova scenario, a crucial question is how frequently ONe novae occur within and around star-forming giant molecular clouds (GMCs) to chemically enrich cold gas. 
Using the results of our recent hydrodynamical simulations of compact gas-rich dwarf galaxies forming GMCs (BM26), we investigate the mass fraction of ONe novae ($F_{\rm ONe}$) within GMCs formed in these galaxies. This fraction $F_{\rm ONe}$ is defined as follows:
\begin{equation}
F_{\rm ONe} = \frac{ M_{\rm ONe} }{ M_{\rm gmc} },
\end{equation}
where $M_{\rm ONe}$ is the total mass of all ONe novae (i.e., binary systems)
interacting with  a GMC.
We assume $M_{\rm ONe}$ to be proportional to the total mass of existing old stars ($M_{\rm s}$) multiplied by the yield of ONe novae per unit star-forming mass ($R_{\rm ONe}$). 
We adopt $R_{\rm ONe}=0.01$ from theoretical models of ONe and CO novae (Kemp et al. 2022) to estimate $F_{\rm ONe}$ for each star-forming GMC in the simulated dwarf galaxies. 
If the distance between an existing star and a GMC is less than $R_{\rm gmc}$, where $R_{\rm gmc}$ is the radius of the GMC, the star is identified as being within the GMC. We adopt the $R_{\rm gmc}-M_{\rm gmc}$ relation from Bekki \& Mackey (2009) to estimate $R_{\rm gmc}$ from $M_{\rm gmc}$.

Fig.~6 shows how $F_{\rm ONe}$ depends on $M_{\rm gmc}$ in the two bulgeless galaxy models (BM26), in which the total masses of dark matter, stars, and gas are $1.7 \times 10^{10}~{\rm M}_{\odot}$, $2.7 \times 10^8~{\rm M}_{\odot}$, and $3 \times 10^7~{\rm M}_{\odot}$, respectively. The gaseous and stellar disk scale lengths are set to 1~kpc and 3~kpc in the two models so that numerous dense GMCs can form within the compact gas disks within $\approx 10^7$~yr.
Clearly, there is a positive correlation between $F_{\rm ONe}$ and $M_{\rm gmc}$, demonstrating that more massive star-forming GMCs $\ge 10^5 {\rm M}_{\odot}$
 can interact with more ONe novae per unit mass.
This means that more massive GMCs are more likely to form P-rich stars due to more efficient P-enrichment of cold gas by ONe novae. 
It is also clear that the number of massive GMCs with non-zero $F_{\rm ONe}$ ($N_{\rm gmc}=807$) is larger in the more compact dwarf galaxy model ($R_{\rm disk}=1$~kpc) than in the model with $R_{\rm disk}=3$~kpc ($N_{\rm gmc}=470$), which implies that P-rich star formation is more efficient in more compact dwarf galaxies.
These results demonstrate that the ONe nova scenario has a robust physical foundation regarding the intense interactions between GMCs and ONe novae.

We cannot accurately predict how many binary systems with massive white dwarfs (leading to ONe novae) interact with a GMC over its lifetime without full hydrodynamical simulations of galaxies including the detailed physics of ONe nova explosions.
However, we can roughly estimate $N_{\rm b}$ from $F_{\rm ONe}$ by assuming a mean mass for binary stellar systems leading to ONe novae ($m_{\rm ONe}$).
For a GMC with $M_{\rm gmc}=10^6~{\rm M}_{\odot}$ 
and $F_{\rm ONe}=10^{-3}$, the total number of binary systems leading to ONe novae,
$N_{\rm b}$, which is $M_{\rm gmc} F_{\rm ONe}/m_{\rm ONe}$, can be $\approx 100$ assuming $m_{\rm ONe}=10~{\rm M}_{\odot}$.
If one massive binary system undergoes $10^4$ ONe nova explosions with each explosion releasing $10^{-5}~{\rm M}_{\odot}$ of gaseous ejecta (JH98), then the total mass of ONe nova ejecta is $\approx 10~{\rm M}_{\odot}$ for $M_{\rm gmc} = 10^6~{\rm M}_{\odot}$. Therefore, the dilution factor in this case is $F_{\rm dil}=10^5$, assuming all gas in the GMC mixes with the ejecta.

Because $M_{\rm ONe}$ around each GMC (and thus $F_{\rm ONe}$) is evaluated at each time step ($\delta t = 1.4 \times 10^6$~yr) in the above estimation, the cumulative $M_{\rm ONe}$ of all ONe novae that a GMC interacts with over its lifetime ($10^7$--$10^8$~yr) could be much larger.
If this is the case, $F_{\rm dil}$ can be as low as $\approx 10^3$. Furthermore, ONe nova ejecta might mix with only a small fraction of pristine GMC gas, which would lower $F_{\rm dil}$ even further. 
Given that the present study predicts different abundance patterns for P-rich stars depending on $F_{\rm dil}$, future work will investigate how many ONe novae interact with GMCs over their lifetimes using more sophisticated hydrodynamical simulations of GMC formation and evolution in 
various galaxy environments in order to assess the validity of
the present scenario in a much more quantitative way.

\begin{figure}
\psfig{file=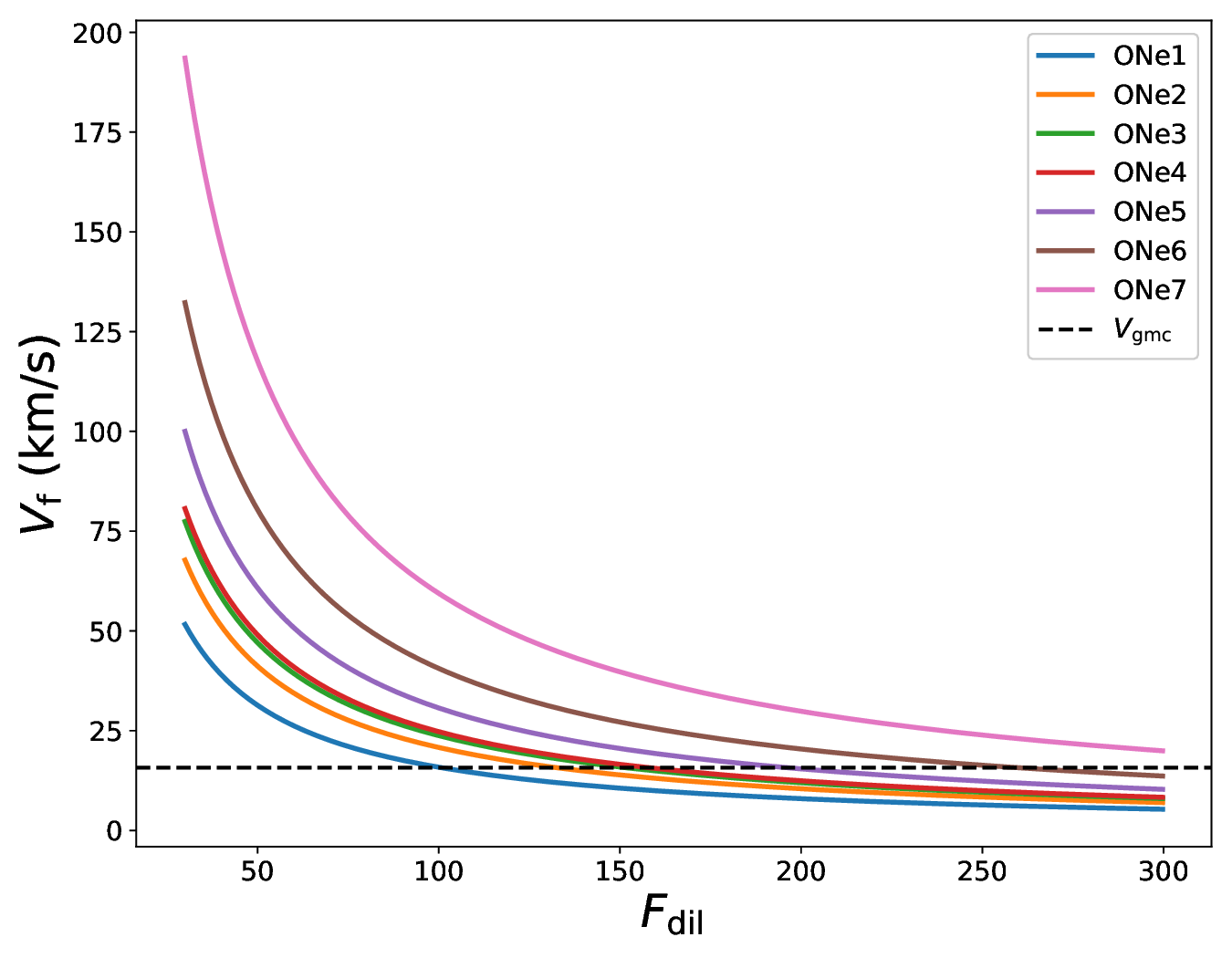,width=8.5cm}
\caption{
Final velocity of the mixed gas, $V_{\rm f}$, as a function of the gas dilution factor $F_{\rm dil}$ for seven ONe nova types (ONe1--7). The horizontal dashed line denotes the escape velocity of a GMC with mass $M_{\rm gmc} = 3 \times 10^6~{\rm M}_{\odot}$ ($V_{\rm gmc} = 15.7~{\rm km~s^{-1}}$). Nova ejecta can be gravitationally trapped by the GMC when $V_{\rm f} < V_{\rm gmc}$.
}
\label{Figure. 7}
\end{figure}

\begin{figure}
\psfig{file=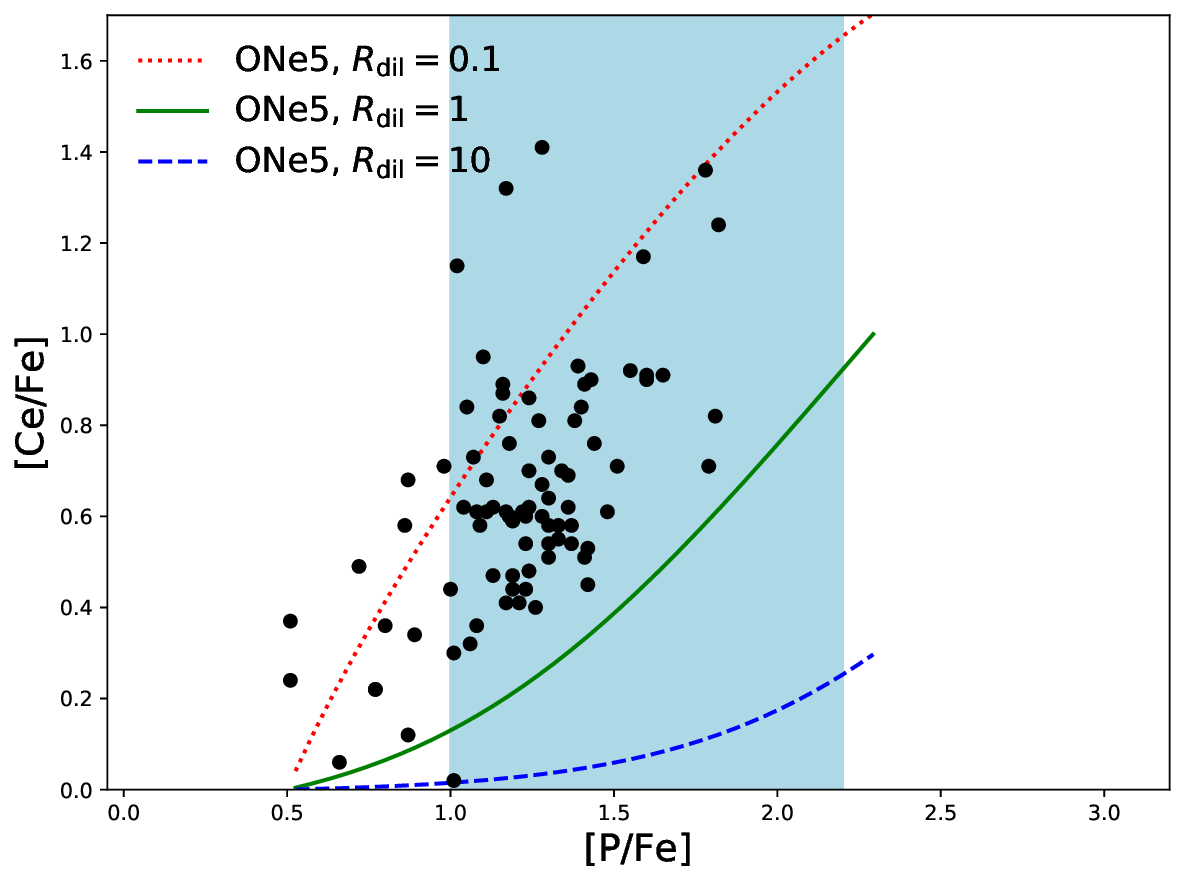,width=8.5cm}
\caption{
 Predicted abundance ratio $[{\rm Ce/Fe}]$ 
as a function of $[{\rm P/Fe}]$ for the three ONe5 models
with $R_{\rm di}=0.1$ (red dotted) 
$R_{\rm di}=1$ (green solid) 
and $R_{\rm di}=10$ (blue dashed).
The vertical light-blue shaded region 
indicates the range of interest for P-rich stars ($1.0 \le [{\rm P/Fe}] \le 2.2$).
The observational data points (black circles) are  from B23.
}
\label{Figure. 8}
\end{figure}

\subsection{Trapping efficiencies of ONe nova ejecta by GMCs}

The ejection velocities of ONe novae ($V_{\rm ej}$) are rather high, ranging from $1600~{\rm km~s^{-1}}$ to $6000~{\rm km~s^{-1}}$ for the ONe1--7 models (JH98). Therefore, the assumption that all ONe nova ejecta are trapped by GMCs might not be entirely realistic. 
Although full hydrodynamical simulations of interactions between cold GMC gas and high-speed ONe nova ejecta are required to assess the validity of this assumption, we investigate this problem using an idealized analytical model. 
We assume that: (i) ONe nova ejecta reach a final velocity ($V_{\rm f}$) after mixing thoroughly with the cold gas of a GMC, (ii) linear momentum is conserved during mixing, and (iii) if $V_{\rm f}$ is less than the escape velocity of the GMC ($V_{\rm gmc}$), the ejecta can be trapped by the GMC. 
Accordingly, $V_{\rm f}$ is a function of $F_{\rm dil}$ and $V_{\rm ej}$:
\begin{equation}
V_{\rm f} = \frac{ V_{\rm ej} }{ (1+F_{\rm dil}) },
\end{equation}
where $V_{\rm f}$ increases for lower $F_{\rm dil}$.
It should be stressed that this $V_{\rm f}$ is likely overestimated in the above equation, as energy dissipation in the shocked gas during the interaction between ONe nova ejecta and GMCs is completely neglected.

Fig.~7 shows whether $V_{\rm f}$ drops below $V_{\rm gmc}$ for the seven ONe nova types assuming $M_{\rm gmc}=3 \times 10^6~{\rm M}_{\odot}$ and $R_{\rm gmc}=103~{\rm pc}$ ($V_{\rm gmc}=15.7~{\rm km~s^{-1}}$).
Clearly, the dilution factor required for trapping ONe nova ejecta ($F_{\rm dil, trap}$) is lower for ONe novae with lower $V_{\rm ej}$ at a given GMC mass: 
for example, $F_{\rm dil, trap} \approx 100$ for ONe1 ($V_{\rm ej}=1600~{\rm km~s^{-1}}$) in the model with $M_{\rm gmc}=3 \times 10^6~{\rm M}_{\odot}$.
Also, ejecta from ONe7 are unlikely to be gravitationally 
trapped by GMCs with $M_{\rm gmc}=3 \times 10^6~{\rm M}_{\odot}$ for $F_{\rm dil} < 300$.
Furthermore, $F_{\rm dil, trap}$ is 
larger for smaller $M_{\rm gmc}$ owing to the lower $V_{\rm gmc}$.

The present study demonstrates that $F_{\rm dil}$ must be relatively low for P-rich star formation with $1 \le [{\rm P/Fe}] \le 2.2$; for example, $F_{\rm dil}$ must be smaller than $\approx 286$ for the single ONe5 model shown in Fig.~1. On the other hand, Fig.~7 indicates that $F_{\rm dil}$ needs to exceed a threshold value ($F_{\rm dil, trap}$) of $196$ for ONe nova ejecta to be gravitationally trapped by a massive GMC. 
Although $F_{\rm dil, th} > F_{\rm dil, trap}$ means that ejecta from ONe5 novae can be trapped efficiently by GMCs with $M_{\rm gmc} = 3 \times 10^6~{\rm M}_{\odot}$, only a narrow range of $F_{\rm dil}$ ($196 \le F_{\rm dil} \le 286$) is permissible for P-rich 
star formation in this simplified argument: this $F_{\rm dil}$ range can be
different between different $M_{\rm gmc}$.
 The allowed range of dilution factors ($F_{\rm dil, trap} \le F_{\rm dil} \le F_{\rm dil, th}$) varies across different ONe nova types.
 Determining whether these required $F_{\rm dil}$ ranges are physically achievable during interactions between GMCs and ONe novae is beyond the scope of this paper. Consequently, numerically investigating the extent to which ONe nova ejecta can be gravitationally retained by GMCs forming massive star clusters will be the subject of future studies.

\begin{figure}
\psfig{file=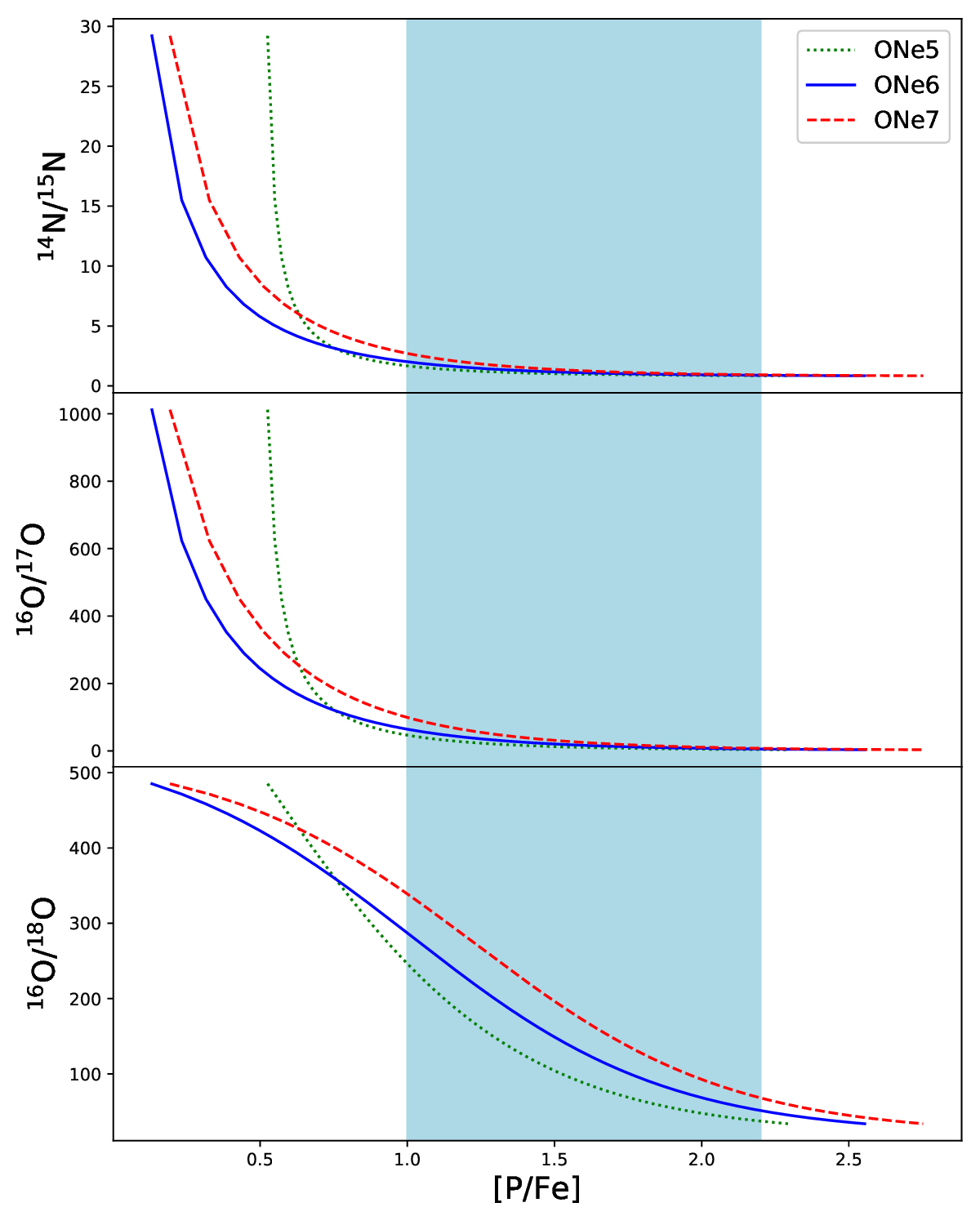,width=8.5cm}
\caption{
 Predicted isotope ratios $^{14}{\rm N}/^{15}{\rm N}$ (top panel), $^{16}{\rm O}/^{17}{\rm O}$ (middle panel), and $^{16}{\rm O}/^{18}{\rm O}$ (bottom panel) as a function of $[{\rm P/Fe}]$ for the ONe5 
(green dotted), ONe6 (blue solid), and ONe7 (red dashed) models.
The vertical light-blue shaded region 
 indicates the range of interest for P-rich stars ($1.0 \le [{\rm P/Fe}] \le 2.2$).
The solar values of
$^{14}{\rm N}/^{15}{\rm N}$, $^{16}{\rm O}/^{17}{\rm O}$, and $^{16}{\rm O}/^{18}{\rm O}$ 
are 441, 2703, and 530, respectively.
}
\label{Figure. 9}
\end{figure}

\subsection{Enhanced abundances of $s$-process elements}

One intriguing discovery in recent observational studies of P-rich stars is that heavy-element abundances (e.g., $[{\rm Ce/Fe}]$ and $[{\rm Ba/Fe}]$) are substantially enhanced (e.g., M20a). 
In the ONe nova scenario, these enhancements result from the joint enrichment of GMCs by both ONe novae and AGB stars. Accordingly, we investigate how $[{\rm Ce/Fe}]$ 
in P-rich stars depends on $R_{\rm dil}$ by adopting the yields predicted for 
low-metallicity AGB stars by 
Fishlock et al. (2014). 
The predicted [Ce/Fe] of AGB stars with their progenitor stellar masses ($m_{\rm s}$) 
of $2.25 {\rm M}_{\odot}$ and $3.0 {\rm M}_{\odot}$ are 2.26 and 1.75, respectively,
in Fishlock et al. (2014), which is consistent with predictions by Lugaro et al. (2012)
for low-metallicity AGB stars.
We assume that: (i) $[{\rm Ce/Fe}] = 2.0$ in the AGB ejecta serves as a reference value, and (ii) P enrichment by these AGB stars can be neglected 
due to their low P yields. 
Fig.~8 demonstrates that P-rich stars with $1.0 \le [{\rm P/Fe}] \le 2.2$ in the ONe5 model 
with $R_{\rm dil} =0.1$ and 1 can reach $[{\rm Ce/Fe}] \approx 0.8$, consistent with observations. 
If larger Ce yields are adopted, the formation of P-rich stars with $[{\rm Ce/Fe}] > 1$ 
becomes possible in the model with $R_{\rm dil}=1$. 
Conversely, the ONe5 model with $R_{\rm dil} = 10$ and
fail to produce P-rich stars with $[{\rm Ce/Fe}] >0.5$, 
implying that a specific range of $R_{\rm dil}$ is required to form P-rich stars with 
high $[{\rm Ce/Fe}]$.
The observed large dispersion in [Ce/Fe] for a given [P/Fe] can be nicely explained,
if GC-forming GMCs have different $R_{\rm dil}$ ranging from 0.1 to 1.0.

Whether such a range of $R_{\rm dil} \approx 0.1-1.0 $ 
is physically feasible in GMCs interacting with ONe novae and AGB stars remains a key question in the ONe nova scenario.
 A single binary stellar system with a massive WD (giving rise to ONe novae) 
can eject $\approx 0.1~{\rm M}_{\odot}$ of gas across $N \approx 10^4$ nova 
outbursts, whereas a low-mass AGB star ($m = 3~{\rm M}_{\odot}$) 
ejects stellar winds amounting to $\approx 2.7~{\rm M}_{\odot}$. 
Accordingly, the expected dilution ratio is $R_{\rm dil} \approx 0.04$ if the 
ejecta from a single ONe nova system and a single AGB star mix with 
GMC gas in equal proportions. 
This estimated value ($R_{\rm dil} \approx 0.04$) is only slightly lower than the required 
range ($R_{\rm dil} \approx 0.1$--$1$), suggesting that the ONe nova scenario 
is promising. 
However, because real GMCs interact with multiple ONe novae and AGB stars, 
the simple framework considering single stellar sources presents an oversimplification.
Since $R_{\rm dil}$
can be more accurately constrained through hydrodynamical simulations of interactions 
between GMCs, AGB stars, and ONe novae, 
future work will investigate the physically 
plausible range of $R_{\rm dil}$ using such numerical models.
Binary systems consisting of intermediate-mass 
stars with masses of $8-10\,\mathrm{M}_{\odot}$ (yielding 
massive white dwarfs) and $3-4\,\mathrm{M}_{\odot}$ (yielding 
low-mass AGB stars) can eject gas through ONe 
novae and AGB phases. If this ejecta mixes with the GMC gas, 
the required $R_{\rm dil}$ condition may be satisfied.

B24 compared the observed heavy-element abundances (e.g., Sr, Ba, and Ag) of P-rich stars 
with predictions from $s$- and $i$-process nucleosynthesis models, finding that a combination of 
$s$- and $i$-process enrichment best reproduces the observed abundance patterns. 
Within the ONe nova scenario, if two distinct types of AGB polluters—one ejecting $s$-process material 
and the other ejecting $i$-process material—interact with growing GMCs, the required dual-process enrichment 
can naturally occur. 
Given that nucleosynthetic models for low-mass and super-AGB stars, including those incorporating 
$i$-process proton ingestion events (e.g., Doherty et al. 2015; Den Hartogh et al. 2019; Choplin et al. 2021), 
do not predict high phosphorus yields ($[\mathrm{P/Fe}] > +1.0$), a combination of ONe novae and 
two distinct AGB polluters (yielding $s$- and $i$-process material, respectively) offers a viable 
framework to explain the heavy-element abundance patterns revealed by B24.

\subsection{Predicted nitrogen and oxygen isotope ratios}

A key prediction from JH98 is that the abundances of nitrogen
and oxygen isotopes in ONe novae are anomalous compared to solar values,
which are 
$^{15}{\rm N}/$$^{14}{\rm N}=2.27 \times 10^{-3}$, 
$^{17}{\rm O}/$$^{16}{\rm O}=3.7 \times 10^{-4}$,
and $^{18}{\rm O}$/$^{16}{\rm O}=2.0 \times 10^{-3}$.
The predicted anomalous N and O isotope  abundances 
imply  that these isotope anomalies could be imprinted on the elemental abundances of 
P-rich stars in the ONe nova scenario.
Furthermore, previous studies on the isotopic patterns of presolar grains 
have found evidence of isotopic anomalies characteristic of ONe novae 
(e.g., Liu et al. 2024), implying that one or more ONe nova explosions occurred within the GMC that hosted the Sun.
We therefore investigate the $^{14}{\rm N}/^{15}{\rm N}$, $^{16}{\rm O}/^{17}{\rm O}$, and $^{16}{\rm O}/^{18}{\rm O}$ isotope ratios in the ONe5, ONe6, and ONe7 models across different values of $F_{\rm dil}$ to provide theoretical predictions for future observations. We limit our investigation to these three models because the other ONe nova models yield excessively high $[{\rm P/Fe}]$ for reasonable ranges of $F_{\rm dil}$.
Fig.~9 illustrates how these isotope ratios depend on $[{\rm P/Fe}]$ in stars for the three representative models. Very low $^{14}{\rm N}/^{15}{\rm N}$ ($<20$) and $^{16}{\rm O}/^{17}{\rm O}$ ($<400$) ratios are clearly evident in P-rich stars with $1.0 \le [{\rm P/Fe}] \le 2.2$ for these  models.

All three models show a qualitatively similar anti-correlation between $[{\rm P/Fe}]$ and the N and O isotope ratios over a wide range of $[{\rm P/Fe}]$. 
This robust prediction can serve to test and validate the ONe nova scenario.
Although the predicted $^{16}{\rm O}/^{18}{\rm O}$ ratio is significantly lower than the solar value, the discrepancy is less extreme than those for $^{14}{\rm N}/^{15}{\rm N}$ and $^{16}{\rm O}/^{17}{\rm O}$.
Gaseous ejecta from AGB stars alter these isotope abundances only slightly because the total mass of AGB ejecta is much smaller than the mass of the GMC.
These results suggest that if such low isotope ratios are detected in main-sequence P-rich stars, it would strongly support the ONe nova scenario, although such measurements for unevolved stars are exceptionally challenging. Nevertheless, measuring these isotope abundances in main-sequence P-rich stars remains a valuable target for future observational studies.

\begin{figure}
\psfig{file=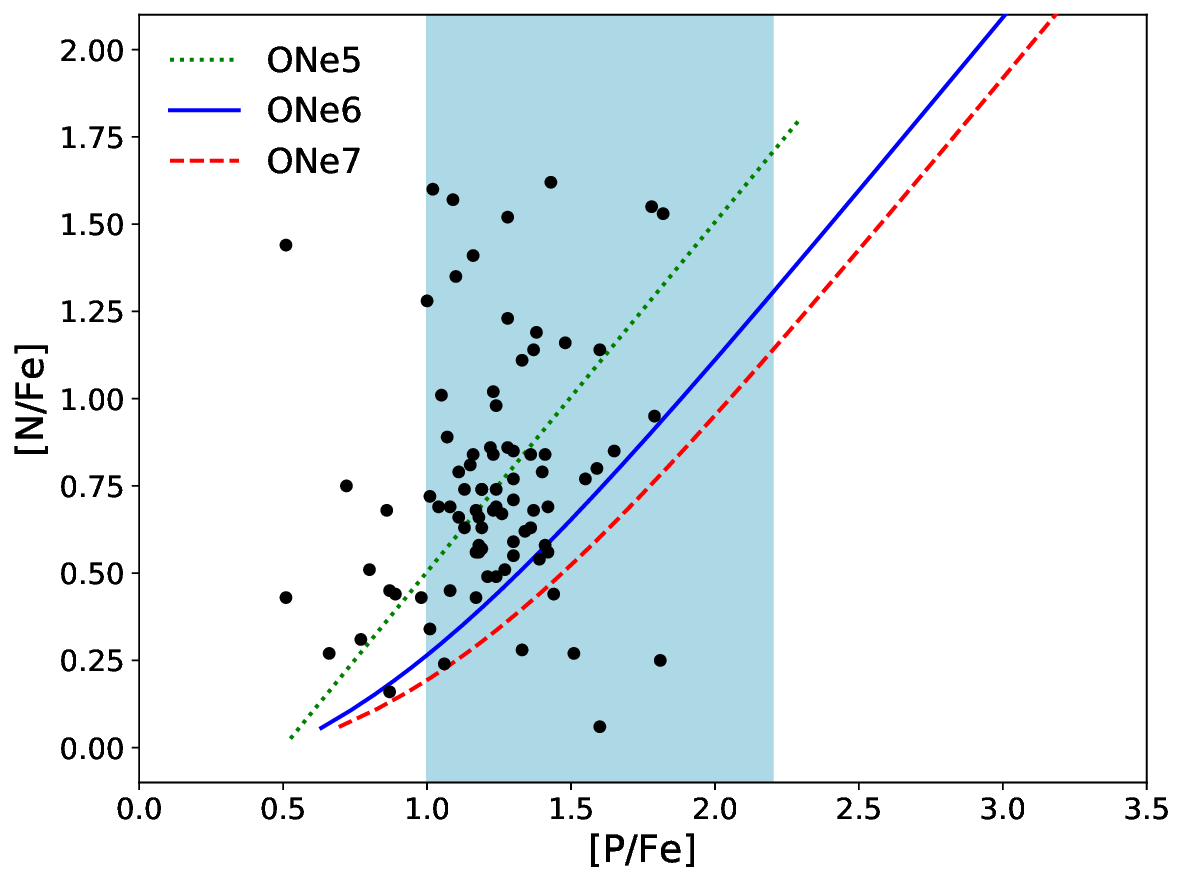,width=8.5cm}
\caption{
Predicted abundance ratio $[{\rm N/Fe}]$ as a function of $[{\rm P/Fe}]$ for the  single
ONe5 (green dotted), ONe6 (blue solid), and ONe7 (red dashed) models.
Black circles are observational data from B23.
}
\label{Figure. 10}
\end{figure}

\subsection{Relations between N-rich and P-rich stars}

The observed high nitrogen abundances ($[{\rm N/Fe}]>0.5$) in 2P stars of Galactic GCs have been successfully reproduced by GC formation scenarios in which massive AGB stars act as the primary polluters (e.g., Bekki et al. 2007; D'Ercole et al. 2008; BM26).
Consequently, P-rich stars can belong to 2P with high $[{\rm P/Fe}]$ and $[{\rm N/Fe}]$ if they form from GMC gas polluted by both massive AGB stars and ONe novae. 
Although a correlation between $[{\rm P/Fe}]$ and $[{\rm N/O}]$ is observed in some GCs (B25), it remains observationally unclear whether P-rich stars are preferentially located in the 2P population of GCs.
Given that nitrogen abundances in ONe novae are quite high (JH98), the observed high $[{\rm N/Fe}]$ in P-rich stars could be due, at least in part, to chemical enrichment by ONe novae.
We thus investigate $[{\rm N/Fe}]$ in P-rich stars formed from GMCs polluted solely 
by ONe novae, assuming: (i) the initial $[{\rm N/Fe}]$ of GMCs is zero (as in 1P stars), 
(ii) only ONe5, ONe6, and ONe7 contribute to P and N enrichment, 
and (iii) $F_{\rm dil}$ ranges from $10$ to $10^4$.

Fig.~10 shows a clear positive correlation between $[{\rm P/Fe}]$ and $[{\rm N/Fe}]$ in all 
three models, which is qualitatively consistent with observations, although the observed dispersion in $[{\rm N/Fe}]$ for a given $[{\rm P/Fe}]$ is large. The ONe5 model shows a more significant 
enhancement ($\Delta[{\rm N/Fe}] > 0.5$) in stars with $1.0 \le [{\rm P/Fe}] \le 2.2$ compared to the ONe6 and ONe7 models. 
Because the initial $[{\rm N/Fe}]$ is set to zero (similar to 1P stars in GCs), these results imply that even 1P stars can become both P-rich and N-rich if they form from GMC gas polluted solely by ONe novae (and not by massive AGB stars). 
Stars formed from gas enriched by both massive AGB stars and 
ONe novae would likewise be classified as P-rich and N-rich 2P stars. 
Conversely, P-normal, N-rich 2P stars can form from gas polluted exclusively by massive AGB stars, implying that not all N-rich stars discovered in the Galaxy (e.g., Schiavon et al. 2017; 
Fern\'andez-Trincado et al. 2020) are necessarily P-rich within this framework.

\begin{figure}
\psfig{file=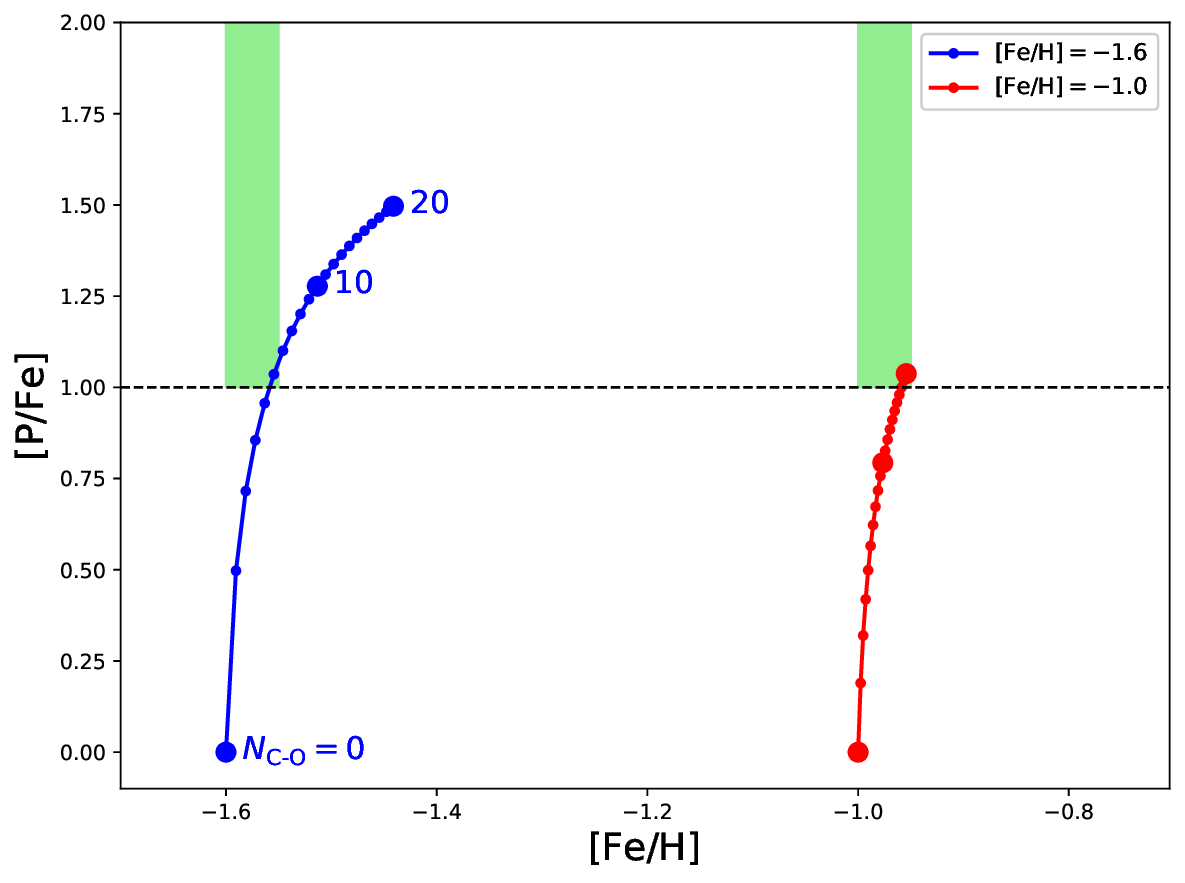,width=8.5cm}
\caption{
Predicted $[{\rm P/Fe}]$ and $[{\rm Fe/H}]$ of GMC gas polluted by CCSNe
with C--O shell merging as a function of the number of CCSNe events ($N_{\rm C-O}$), 
shown for initial metallicities $[{\rm Fe/H}] = -1.6$ and $-1.0$.
The vertical light-green shaded bands around [Fe/H]$=-1.6$ and $-1$ denote the observed ranges of
[Fe/H] and [P/Fe] of P-rich stars in Type I GCs with [Fe/H] spreads of less than 0.05 dex.
}
\label{Figure. 11}
\end{figure}

\begin{figure*}
\psfig{file=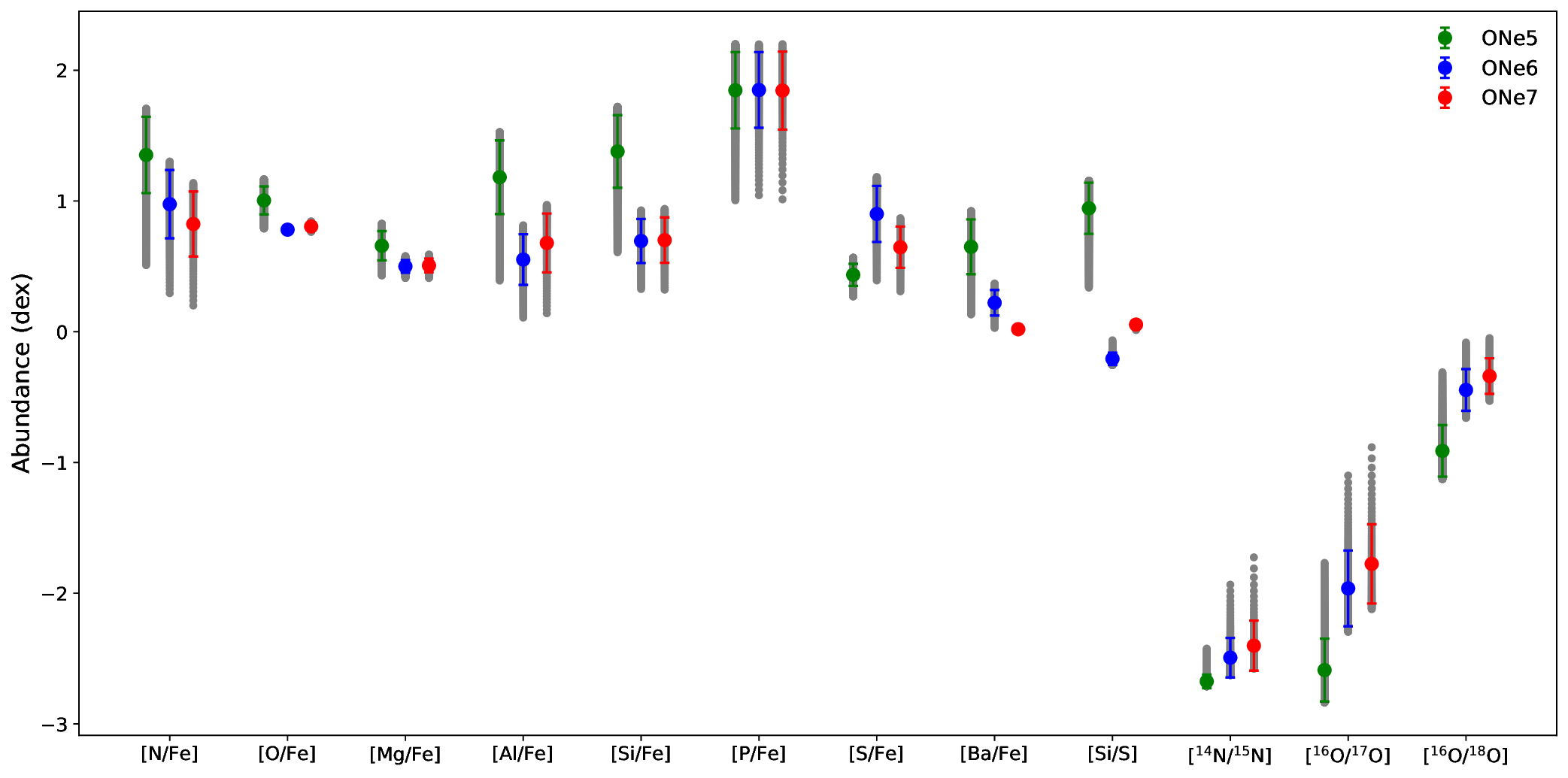,width=18cm}
\caption{
Predicted mean values and $1\sigma$ dispersions of elemental and isotopic abundance ratios ($[{\rm N/Fe}]$, $[{\rm O/Fe}]$, $[{\rm Mg/Fe}]$, $[{\rm Al/Fe}]$, $[{\rm Si/Fe}]$, $[{\rm P/Fe}]$, $[{\rm S/Fe}]$, $[{\rm Ba/Fe}]$, $[{\rm Si/S}]$, $[^{14}{\rm N}/^{15}{\rm N}]$, $[^{16}{\rm O}/^{17}{\rm O}]$, and $[^{16}{\rm O}/^{18}{\rm O}$, in dex) for P-rich stars, derived from the single ONe5 (green), 
ONe6 (blue), and ONe7 (red) models. The vertical gray bars represent the underlying distribution of values across the range of dilution factors.
}
\label{Figure. 12}
\end{figure*}

\subsection{A possible explanation for the observed [Fe/H] range of P-rich stars}

Although the observed $[\mathrm{Fe/H}]$ distribution of P-rich stars overlaps significantly 
with that of Galactic GCs (M20a), two major differences exist between their metallicity distributions. 
First, P-rich stars with $[\mathrm{Fe/H}] < -1$ are relatively rare, 
whereas Galactic GCs exhibit a prominent 
peak around $[\mathrm{Fe/H}] = -1.6$. 
Second, P-rich stars with $[\mathrm{Fe/H}] > -0.5$ have not yet been observed, whereas the Galactic bulge 
hosts GCs with metallicities exceeding $[\mathrm{Fe/H}] = -0.5$. 
These two observational facts have yet to be fully explained by current formation scenarios for P-rich stars, 
though they provide strong constraints on the nucleosynthetic sites of phosphorus enrichment.

The observed scarcity of metal-poor P-rich stars ($[\mathrm{Fe/H}] < -1$) can be explained within the ONe nova scenario as follows. 
Using binary population synthesis models, Kemp et al. (2022) showed that (i) novae begin exploding 
$\approx 10^8~\mathrm{yr}$ after the onset of star formation, and (ii) their event rates peak around 
$3 \times 10^8$ and $3 \times 10^9~\mathrm{yr}$ after star formation begins (see their Fig.~11 for $Z = 0.0001$). 
These results imply that GC-forming GMCs rarely interact with ONe novae during the very early phases of galaxy 
formation when $[\mathrm{Fe/H}] < -1$, although we note that their calculations represent combined CO and ONe novae. 
Furthermore, the nova event rate drops substantially in later galactic evolutionary stages when $[\mathrm{Fe/H}] > -0.5$ (Kemp et al. 2022), 
which may also account for why P-rich stars with $[\mathrm{Fe/H}] > -0.5$ have not been discovered.

\subsection{Contributions of CCSNe with C-O shell merging to P-enrichment}

Using 3D hydrodynamical simulations of massive stars undergoing C-O shell merging, R18 demonstrated that the 
overproduction factor of P (the ratio of final to initial P abundances) can reach values as high as 10 in models with $Z=0.01$ and $0.02$.
If these massive stars eject all of their newly synthesized 
P through subsequent CCSNe and if the ejecta mix thoroughly with cold GMC gas, P-rich stars could potentially form within GMCs.
However, CCSNe also enrich GMCs in Fe, which causes newly formed stars to exhibit a broader spread in $[{\rm Fe/H}]$. Accordingly, the small observed $[{\rm Fe/H}]$ spreads ($<0.05$~dex) in Type I GCs containing P-rich stars (e.g., M4) impose stringent constraints on the contribution of CCSNe to chemical enrichment during GC formation.

Here, we investigate $[{\rm Fe/H}]$ and $[{\rm P/Fe}]$ in stars formed from cold GMC gas polluted by CCSNe with C-O shell mergers for initial $[{\rm Fe/H}] = -1.6$ and $-1.0$, assuming $[{\rm P/Fe}] = 2.0$ in the ejecta (corresponding to the maximum possible P abundance in R18).
As shown in Fig.~11, $[{\rm P/Fe}]$ can exceed 1.0 
if the umber of CCSNe with C-O shell merging ($N_{\rm C-O}$) is larger than 6
 at $[{\rm Fe/H}]=-1.6$; however, $[{\rm Fe/H}]$ also increases beyond $-1.65$, introducing an $[{\rm Fe/H}]$ spread that exceeds observed limits. 
Although $[{\rm P/Fe}]$ reaches 1.0 in the model with $N_{\rm C-O}=20$ and $[{\rm Fe/H}]=-1.0$, $[{\rm Fe/H}]$ increases to values above $-0.95$. If $N_{\rm C-O}>20$, $[{\rm P/Fe}]$ can be significantly higher than 1.0 at $[{\rm Fe/H}]=-1.0$, but the resulting $[{\rm Fe/H}]$ spreads exceed $0.05$~dex, which is inconsistent with observations.
Note that only a small fraction of CCSNe undergo C-O shell merging (e.g., $\approx 20$\% in Roberti et al. 2025). The combined contribution of CCSNe with and without C-O shell merging would produce even larger $[{\rm Fe/H}]$ spreads in GC stars than estimated above.

These results imply that P-rich stars in Type I GCs are highly unlikely to form in GMCs 
polluted by CCSNe with C-O shell merging. 
Furthermore, R18 predicted significant K and Sc enhancements in the ejecta of massive 
stars with C-O shell merging, whereas ONe nova models show no such enhancement (e.g., JH98). 
Barbuy et al. (2025) investigated the chemical abundances of P, S, and K in 58 Galactic bulge 
stars and found that P-rich stars do not show significant K enhancement. 
This result is consistent with the ONe nova scenario 
rather than the C-O shell merger scenario. 
Assuming that all field P-rich stars originate from GCs, 
measuring [K/Fe] and [Sc/Fe] spreads in GCs containing P-rich stars 
provides an additional observational test to distinguish between these two scenarios.

\subsection{Future work}

The present study has demonstrated that the single ONe5 models best reproduce 
the observed abundance patterns of P-rich stars with a specific range of $F_{\rm dil}$
among single ONe nova models.
The composite models can partly reproduce the observed ranges of ${\rm [\alpha/Fe]}$, [Al/Fe],
and [Si/S] for a given [P/Fe] with different $F_{\rm dil}$,
they cannot so well the abundance patterns of some P-rich stars.
Fig.~12 summarizes the predicted abundance ranges of P-rich stars 
with $1.0 \le [{\rm P/Fe}] \le 2.2$ for various elements in the ONe5, ONe6, and ONe7 models
with different $F_{\rm dil}$
as reference models.
Because $F_{\rm dil}$ is treated as a free parameter in this study, 
it remains theoretically unclear whether the required dilution factor of $F_{\rm dil}$ for each
model  is physically attainable in GC-forming GMCs.

Therefore, a primary objective of future work is to understand what determines $F_{\rm dil}$ when ONe nova ejecta interact with cold GMC gas. 
Future hydrodynamical simulations of GC formation in GMCs polluted by ONe novae will investigate what fraction of the high-velocity ($>1000~{\rm km~s^{-1}}$) and high-temperature ($T>10^8~{\rm K}$) ejecta can be retained within GMCs after mixing with cold gas. The formation of P-rich stars may be possible from mixed gas at high densities ($>10^5~{\rm cm^{-3}}$) if the gas temperature drops rapidly to $\sim 10~{\rm K}$. 
Because GCs are observed to exhibit O--Na, C--N, and Mg--Al anticorrelations, the enhanced N, O, and Mg abundances resulting from ONe nova enrichment could weaken or erase these anticorrelations if enrichment occurs throughout the entire GMC.
It is therefore also important for future work to investigate what mass fraction of a GMC is polluted by ONe novae and what fraction of stars in GCs become P-rich.

Future observational studies measuring $[{\rm P/Fe}]$ in metal-poor GCs ($[{\rm Fe/H}] < -1.2$) are essential to determine whether P-rich stars are ubiquitous across GCs of 
different metallicities, given that previous discoveries were confined to GCs 
with $[{\rm Fe/H}] \ge -1.2$ (e.g., M4).
If future observations confirm the absence of P-rich stars with $\mathrm{[Fe/H]} < -1.2$,
the ONe nova scenario can account for these findings,
as discussed in \S 4.6. Conversely, if P-rich
stars with $\mathrm{[Fe/H]} < -1.2$ are discovered,
their observed minimum $\mathrm{[Fe/H]}$ can be used to
constrain whether chemical enrichment by ONe novae can
occur within GC-forming GMCs while their host dwarf galaxies
are still in their earliest formation phases.

The dependence of $[\mathrm{P/Fe}]$ on $[\mathrm{Fe/H}]$ in Galactic stars 
has been investigated by several groups and 
compared with the $[\mathrm{Fe/H}]$ trends of other elements (e.g., Caffau et al. 2019; 
Maas et al. 2021; Nandakumar et al. 2022;  Esposito et al. 2026; Gakis et al. 2026). 
This relation is strikingly different from that of $[\alpha/\mathrm{Fe}]$ at $[\mathrm{Fe/H}] < -1$ (see, e.g., Fig. 1 in Andrievsky \& Korotin 2026 for recent results), 
where a population of P-normal stars ($[\mathrm{P/Fe}] < 0.5$) also exists.
The ONe nova scenario offers a compelling explanation 
for the formation of these P-normal stars. 
Although ONe nova explosions can occur within the low-density diffuse interstellar medium (ISM), 
their ejecta cannot be efficiently trapped by or mixed with the diffuse gas 
owing to its low density. 
Consequently, the resulting mixed gas not only retains a lower $[\mathrm{P/Fe}]$ 
ratio but also remains at densities too low to trigger 
star formation. 
This enriched gas may instead be accreted onto star-forming 
gas clouds much later, eventually forming P-normal stars. 
Accordingly, the ONe nova scenario implies that the local environment 
surrounding ONe novae dictates whether stars formed from ONe-polluted gas become P-rich or P-normal. Future studies must therefore evaluate the trapping efficiency of ONe nova ejecta across ISM environments with diverse physical properties (e.g., gas density).

The physical origin of Type~II GCs exhibiting broad metallicity spreads 
remains one of the key unresolved problems in GC formation. 
Bekki \& Tsujimoto (2016) proposed that the merging of Type~I GCs formed at different 
locations and times within dwarf galaxies is the 
primary mechanism for Type~II GC formation. 
This merger scenario implies that if Type~I GCs universally 
host P-rich stars, Type~II GCs should also contain P-rich stellar populations. 
Although previous observations have determined the chemical abundances of various elements 
in Type~II GCs such as NGC~1851 (e.g., Yong \& Grundahl 2008; Lardo et al. 2012), 
M22 (e.g., Marino et al 2009; McKenzie et al. 2022), 
and $\omega$~Cen (e.g., Johnson \& Pilachowski 2010;
Akbaba et al. 2026),
P-rich stars have not yet been detected in these clusters. 
If future observations reveal P-rich stars spanning 
different $\mathrm{[Fe/H]}$ values in Type~II GCs, such a 
discovery would strongly support the GC merger scenario. 
We plan to investigate the physical properties of P-rich stars in merged Type~II GCs to provide predictions for comparison with future observational data.

Matsuno et al. (2025) recently discovered a Li-rich, metal-poor star, D25\_6334, 
which shows no evidence of a binary companion (i.e., highly likely a single star) 
and exhibits chemical abundance patterns consistent with pollution by ONe6 novae. 
Because ONe6 novae can eject significant amounts of both Li and P (JH98),
this Li-rich star is very likely P-rich as well. 
Although it may have initially formed in a binary or triple system, 
as proposed by Matsuno et al. (2025), 
it is also possible that the star formed as a single star 
within a GMC polluted by ONe novae, as described by our ONe nova scenario. 
If single stars in the Galactic halo are confirmed to be both Li-rich and P-rich, 
their Li abundances will need to be investigated within the context of the 
ONe nova scenario in future work. 
Furthermore, as ongoing near-infrared surveys of Galactic stars Matsunaga et al. (2026)
 uncover new P-rich stars and measure their detailed elemental abundances, our future studies will aim to constrain their origins within this theoretical framework.

\section{Conclusions}

Using analytical models of GC formation within growing GMCs, 
we investigate the chemical abundances of various elements in phosphorus-rich (P-rich) 
stars ($[\mathrm{P/Fe}] \ge 1$) under the ``ONe nova'' scenario. 
In this framework, P-rich stars can form within growing GMCs when ONe nova explosions chemically 
enrich the cold, dense molecular gas. 
Efficient star formation from ONe nova ejecta mixed with cold gas is facilitated by 
hydrodynamic interactions: high-velocity ejecta ($> 1000~\mathrm{km~s^{-1}}$) can be trapped 
and decelerated by the dense ambient GMC gas. 
This ONe nova mechanism is consistent with the ``star--cloud interaction'' scenario 
for GC formation featuring multiple stellar populations (BM26), wherein GC-forming GMCs interact 
with pre-existing stellar populations, such as AGB stars, to form both 
first-population (1P) and second-population (2P) stars. 
Consequently, GCs with multiple stellar populations host P-rich stars only if their progenitor GMCs 
interacted with pre-existing ONe novae; thus, not all GCs are expected to contain P-rich stars in 
the ONe nova scenario.

We have adopted theoretical chemical yields of various elements from JH98 to predict 
the chemical abundances of P-rich stars across a wide range of gas dilution 
factors ($F_{\rm dil}$) for seven ONe nova types (ONe1--7). We have investigated both ``single'' models 
with only one ONe nova type (e.g., ONe1) and ``composite'' models with multiple 
ONe nova types. 
We focus primarily on models with an initial GMC gas metallicity of $[\mathrm{Fe/H}] = -1$, 
as P-rich stars with $[\mathrm{Fe/H}] < -1.2$ are rare in the Galaxy. 
We employ simple analytical models for the interaction between GMCs and ONe nova ejecta, 
alongside hydrodynamic simulations of GMC formation within dwarf galaxies, to evaluate 
the retention probability of ONe nova ejecta within GMCs and the frequency of interactions 
between growing GMCs and ONe novae. 
Our principal results are as follows: \\

(1)  P-rich stars with $[{\rm P/Fe}] \ge 1$ can form in models with different ONe nova types 
(e.g., ONe5) across a wide range of $F_{\rm dil}$. The threshold $F_{\rm dil}$ below which $[{\rm P/Fe}]$ of stars exceeds 1 in single models depends strongly on the adopted ONe type. 
For example, models with ONe5 show $[{\rm P/Fe}] \ge 1$ for $F_{\rm dil} \le 286$ (
i.e., $F_{\rm dil, th}=286$) and $A(P)=A_{\rm JH}(P)$, 
whereas those with One7 reach such high [P/Fe] for $F_{\rm dil} \le 2500$.
This large difference in threshold $F_{\rm dil}$ is due to vastly different P mass fractions in ejecta among ONe nova types, 
ranging from $1.3 \times 10^{-3}$ (ONe5) to $1.2 \times 10^{-2}$ (ONe7). 
Composite models also show P-rich star formation across a wide range of $F_{\rm dil}$. 

(2) The $[\mathrm{P}/\mathrm{Fe}]$ ratio in stars correlates positively with $[\mathrm{O}/\mathrm{Fe}]$, $[\mathrm{Mg}/\mathrm{Fe}]$, $[\mathrm{Al}/\mathrm{Fe}]$, and $[\mathrm{Si}/\mathrm{Fe}]$ in the present models. Specifically, P-rich stars with $1 \le [\mathrm{P}/\mathrm{Fe}] \le 2.2$ exhibit significant enhancements in $[\mathrm{O}/\mathrm{Fe}]$, $[\mathrm{Mg}/\mathrm{Fe}]$, $[\mathrm{Al}/\mathrm{Fe}]$, and $[\mathrm{Si}/\mathrm{Fe}]$ across most single ONe nova models. For the adopted yields, the observed abundances of these elements are best reproduced by the ONe5 models, suggesting that ONe5 novae are the primary polluters responsible for the observed abundance patterns in P-rich stars. \\

(3) The predicted positive correlations of $[{\rm P/Fe}]$ with $[{\rm Si/Fe}]$ and $[{\rm Si/S}]$ are consistent with observations of P-rich stars, and these correlations are not predicted in theoretical models of P-rich star formation based on other polluters such as 
CCSNe with C--O shell merging. It is thus suggested that these positive correlations can be used to distinguish between different theoretical models of P-rich star formation in GCs. \\

(4) The present models predict unusually low $^{14}{\rm N}/^{15}{\rm N}$, $^{16}{\rm O}/^{17}{\rm O}$, and $^{16}{\rm O}/^{18}{\rm O}$ isotope ratios compared to solar values, which are unique to the ONe nova scenario. In particular, extremely low $^{14}{\rm N}/^{15}{\rm N}$ ratios can be seen in the models, irrespective of the adopted ONe nova types. We therefore suggest that future observational measurements of these isotopes for main-sequence stars can be used to test or validate the scenario, although such observational measurements remain technically challenging. \\

(5)  P-rich stars with larger $[{\rm P/Fe}]$ can have larger $[{\rm N/Fe}]$ as well in the
present model, because ONe novae eject N-rich and P-rich gas. Because N-rich (2P) 
stars can form in GC-forming GMCs polluted by massive AGB stars with $m = [4-8] {\rm M}_{\odot}$
 without pollution by ONe novae, N-rich stars are not necessarily P-rich ones. If ONe novae do not occur in GC formation at all, 2P stars in GCs will be N-rich but not P-rich. On the other hand, if ONe novae occur before chemical enrichment by 
massive AGB stars, even 1P stars can be P-rich in the present ONe nova scenario. Furthermore, 
P-rich field stars can originate from dissolved GCs containing P-rich stars. \\

(6) If GC-forming GMCs are co-polluted by ONe novae and pre-existing low-mass AGB stars (which eject Ce- and Ba-rich gas), 
the resulting P-rich stars will display 
enhanced $[\mathrm{Ce}/\mathrm{Fe}]$ and $[\mathrm{Ba}/\mathrm{Fe}]$. 
Accordingly, this scenario provides a possible
 explanation for the observed $s$-process element 
enhancements among P-rich stars. However, whether 
such synchronized pollution of GMCs by both ONe novae and AGB stars 
is physically feasible remains a critical 
open question. This warrants detailed investigation in future work 
using hydrodynamical simulations of GC formation within growing GMCs. \\

(7) The spatial number density of ONe novae within and around GMCs ($\Sigma_{\rm ONe}$) determines whether P-rich stars can form in the ONe nova scenario. Accordingly, star clusters forming in present-day galaxies with relatively low $\Sigma_{\rm ONe}$ are unlikely to produce P-rich stars in this scenario. Ideal environments for P-rich star formation in the scenario are compact, gas-rich 
dwarf galaxies at high redshift, where high-density GMCs surrounded by ONe novae can form. \\

(8) Although P-rich stars can form from GMC gas polluted by CCSNe with C--O shell merging, such enrichment by CCSNe inevitably introduces a large $[{\rm Fe/H}]$ spread 
among GC stars, which is inconsistent with the small $[{\rm Fe/H}]$ 
spreads observed in Type I GCs. Accordingly, CCSNe with 
C--O shell merging would be
  unlikely to be the major polluter for P-rich star formation. 
The C--O shell merger model predicts enhanced K and Sc abundances in P-rich stars, 
which can be used to distinguish it 
from the ONe nova scenario, where no such enhancement is predicted. \\

\section{DATA AVAILABILITY}
The data used in this paper (outputs from computer simulations)
will be shared on reasonable request
to the corresponding author.

\section{Acknowledgment}
KB and TT  would like to acknowledge the referee   Beatriz Barbuy
for thoughtful and detailed comments.
The hydrodynamical simulations of gas-rich disks galaxies
were  performed on the OzSTAR national facility at Swinburne University of Technology.
The OzSTAR program receives funding in part from the Astronomy National
Collaborative Research Infrastructure Strategy (NCRIS) allocation provided by the
Australian Government, and from the Victorian Higher Education State Investment
Fund (VHESIF) provided by the Victorian Government.
 TT acknowledges the support from JSPS KAKENHI Grant No. 26K22391. 

{}

\appendix

\section{Illustration of the scenario}

The ONe nova scenario for P-rich star formation in the present study is 
based on the SCI scenario (BM26), in which globular clusters (GCs) 
form within growing giant molecular clouds (GMCs) polluted by pre-existing stellar populations in dwarf galaxies, such as AGB stars, ONe novae, and core-collapse supernovae (CCSNe; 
see Fig.~A1 for an illustration). In this scenario, high-velocity gas ($>1000~\mathrm{km~s^{-1}}$) ejected by ONe novae is efficiently trapped and mixed with the high-density cold 
gas of GC-forming GMCs, eventually forming new stars. If the dilution factor is 
low ($F_{\mathrm{dil}} \approx 100$), these newly formed stars 
attain $[\mathrm{P}/\mathrm{Fe}] \ge 1$ and are thus identified as P-rich stars. 
If GCs harboring P-rich stars undergo dynamical disintegration, 
these stars are released into the Galactic field. 
The scenario thus implies that field P-rich stars originate 
primarily from low-mass GCs that are especially prone to 
tidal disruption and dynamical disintegration.

A high cold-gas density ($\sim 10^5~\mathrm{cm^{-3}}$) is essential for 
trapping the ONe nova ejecta. If the mixed gas is additionally enriched by pre-existing low-mass AGB stars, which eject Ce- and Ba-rich material, the resulting stars will exhibit 
simultaneously elevated $[\mathrm{P}/\mathrm{Fe}]$, $[\mathrm{Ce}/\mathrm{Fe}]$, and $[\mathrm{Ba}/\mathrm{Fe}]$, as observed in P-rich stars. Because the present work relies on idealized 
analytic models, detailed numerical studies of these key physical processes—such as 
ejecta mixing, star formation efficiency, and 
AGB pollution—will be required in future work to confirm the feasibility of this proposed scenario.
We will investigate these physical processes related to P-rich star formation in 
galaxy-scale simulations using an updated version of our previous GC formation model
(Bekki et al. 2002).

\begin{figure}
\psfig{file=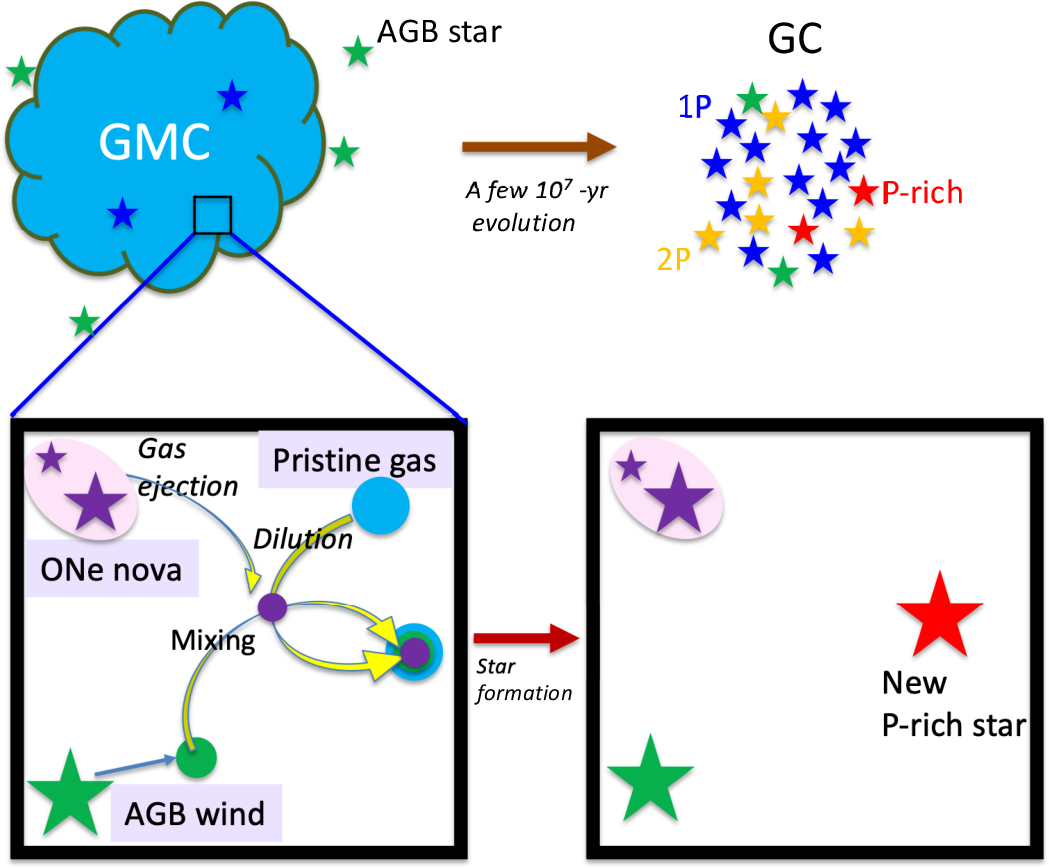,width=8.5cm}
\caption{
An illustration of the ONe nova scenario. In this scenario, a giant molecular cloud (GMC) grows from smaller seed molecular clouds (MCs) via mergers and gas accretion. During its growth, the 
forming GMC interacts with pre-existing stars in the host dwarf galaxy, such as AGB stars, ONe novae, and supernovae. If new stars form from cold molecular gas polluted by ONe novae with a 
relatively small dilution factor ($F_{\mathrm{dil}} \approx 100$), they can become P-rich stars. If the cold gas is additionally enriched by AGB stars, the resulting P-rich stars will also display elevated $[\mathrm{Ce}/\mathrm{Fe}]$ and $[\mathrm{Ba}/\mathrm{Fe}]$. Furthermore, star formation within the GMC can yield a globular cluster (GC) hosting multiple stellar populations. The physical processes underlying GC formation with multiple stellar populations are detailed in BM26.
}
\label{Figure. 13}
\end{figure}

\end{document}